\documentclass[pre,twocolumn,floatfix,reprint]{revtex4}
\usepackage{graphicx,amssymb,amsmath}
\usepackage{stmaryrd}
\usepackage{hyperref}

\begin{document}

\title{Modeling cities}

\author{Marc Barthelemy}
\email{marc.barthelemy@ipht.fr}
\affiliation{Institut de Physique Th\'{e}orique, CEA, CNRS-URA 2306, F-91191, 
Gif-sur-Yvette, France}
\affiliation{CAMS (CNRS/EHESS) 190-198, avenue de France, 75244 Paris Cedex 13, France}

\begin{abstract}

  Cities are systems with a large number of constituents and agents
  interacting with each other and can be considered as emblematic of
  complex systems. Modeling these systems is a real challenge and
  triggered the interest of many disciplines such as quantitative
  geography, spatial economics, geomatics and urbanism, and more
  recently physics. (Statistical) Physics plays a major role by
  bringing tools and concepts able to bridge theory and empirical
  results, and we will illustrate this on some fundamental aspects of
  cities: the growth of their surface area and their population,
  their spatial organization, and the spatial distribution of
  activities. We will present state-of-the-art results and models but
  also open problems for which we still have a partial understanding
  and where physics approaches could be particularly helpful. We will
  end this short review with a discussion about the possibility of
  constructing a science of cities.

\end{abstract}


\maketitle

\section{Introduction}

Modeling cities and getting reliable quantitative predictions about
their behavior have become major challenges for our modern
world. Concentration of humans in a relatively small spatial area
is essentially the result of economical considerations but spatial
localization also leads to an increase of housing costs, traffic
congestion, urban sprawl, pollution and other environmental
problems. Organizing the city, building new infrastructures and new
transportation means rely then a lot on our understanding of their
effects on the collective behavior and their potential
impact on the city, its functioning and its economical
activity. This is particularly important at a time when we expect
the number of megacities (with population larger than 10 millions) to
be around 30 in 2050 \cite{UN:2018} (see also \cite{Barthelemy:2016}
for a more thorough discussion), and emergent countries see their
cities exploding in size: it is estimated that by 2050 the world
population will add 2.5 billion people to urban areas, with $90\%$ of
this increase that will take place in Africa and Asia \cite{UN:2018}.

From a theoretical point of view, cities comprise all the difficulties
inherent to complex systems: a large variety of agents interacting
with each other, a large variety of temporal and spatial scales, and a large number of
processes that modify the spatial structure and organization of these
systems. The always growing availability of urban data will hopefully
help us to construct robust models and thereby to improve our
knowledge of the main mechanisms at play during the evolution of urban
systems. This fundamental level is necessary in order to construct
robust computational models for describing more realistic or specific
situations. Indeed, simulating the behavior of a urban system by
`simply' adding together all the possible processes and gathering as
much information as possible, is neither a guarantee of its
robustness, nor of the correctness of its predictions. 

Economics and regional science were the first disciplines to tackle
this problem of modeling cities. Most of the urban economics is built
on the model proposed in the 60s by Alonso, Muth and Mills (see for
example the textbook \cite{Fujita:1989} and references therein). This
model relies on different assumptions. The first important one is that
all individuals behave in the same way and maximize the same utility
function subject to the same budget constraint. The second assumption
is the monocentric city structure organized around a unique central
business district (CBD). The information about transport
infrastructure is here absent and space is considered as homogeneous
and isotropic. All individuals are assumed to work at this CBD and the
transportation cost depends on the distance from home to it (here we
can see the impact of the first model proposed by Von Thunen
\cite{Vonthunen:1966} with a unique market in an isolated state, where
space is homogeneous and isotropic, and transportation cost depend
only on the euclidean distance to the central market). Later, Fujita
and Ogawa \cite{Fujita:1982} considered a general model where both
individuals and firms are finding their optimal location in the
city. In particular they introduced an agglomeration effect that
describes the benefit for companies to be close to each other. These
approaches constitute the basis of urban economics and we refer the
interested reader to the books \cite{Fujita:1989,Fujita:1999}.

Agent-based models and large simulations also played an important role
for understanding cities and were primarily developped by geographers in
the 50-60s. The primary goal was to forecast the effects of planning
policities and several formalisms were used
\cite{Batty:2008c,Pumain:2013}. The complexity of these simulations,
the large number of parameters and processes however make it difficult
to test thoroughly these models. In addition to these numerical
simulations, it thus seems important to develop parsimonious models
that help identifying the main mechanisms and to construct robust
models. In this respect physics and more precisely statistical physics
with its ability to connect microscopic behaviors to the emergence of
collective and non-trivial behavior seems to be a particularly
well-suited tool. In fact, statistical physics came into play later
and connections with cities appeared first in the context of fractals
and percolation. The fractal dimension of city boundaries was measured
\cite{Batty:1994,Tannier:2005} with a value between $1.2$ and
$1.4$. The diffusion limited aggregation (DLA) model
\cite{Witten:1983} was then naturally invoked to describe the growth
of such a fractal structure. Later, statistical physicists
\cite{Makse:1995} proposed an alternative percolation model where the
presence of correlations between new units is able to predict results
in agreement with the dynamics of cities. This correlated percolation
model \cite{Makse:1998} is a very simplified model for cities, but
nevertheless suggested the possible relevance of simple statistical
physics approaches for systems as complex as cities. The importance of
percolation in cities was reinforced in the study
\cite{Rozenfeld:2008} where the authors proposed a percolation-like
algorithm to define cities without ambiguities, as the giant
percolation cluster of the built-up area. Another important question
where statistical physics contributed a lot is the social structure of
cities. First discussed by Schelling \cite{Schelling:1971}, agents are
described by Ising like spins with tolerance level for other kinds and
was later naturally rediscussed and improved by physicists
\cite{Vinkovic:2006,Grauwin:2009,Gauvin:2009,Dallasta:2008,Jensen:2018}.

In this paper we won't address all aspects of cities but we will focus
on few of the most recent approaches that combine concepts and
ingredients from economics and geography with tools and ideas from
statistical physics. We will first discuss in sections II and III the -- still relatively open
-- problem of the population and area growth of cities and their quantitative
characterization. In particular, we will discuss a candidate model in
direct connection with important problems in statistical physics for
explaining the Zipf law that governs the statistics of urban
populations. In section IV we will consider cities at a intra-urban
level and discuss their spatial structure and the distribution of
activities. Finally, in sections V and VI, we will discuss the issue
about mixing together data for different cities and the possibility of
constructing a `science of cities' \cite{Batty:2013}.

\section{Urban sprawl}

A first and natural question about cities concerns their spatial
extent and how it evolves in time. There are several definitions of
cities but we will assume here a `functional' definition where the
city is not limited to its central core but includes peripheral areas
that exchange enough commuters with it (for example MSA in the US or EU-OECD
functional urban areas in Europe, see for example
\cite{Bettencourt:2013b}). Another possibility is to consider the
percolating cluster of built-up areas as in the `City clustering
algorithm' (CCA) discussed in \cite{Rozenfeld:2008}. 

Cities and their built-up area are usually growing with time and the impact of urban sprawl on the
environment has largely been debated and is highly politicized. There
is however a relatively fair consensus about its negative impact on the
environment (see for example \cite{Brueckner:2000} and references
therein) with the reduction of green areas, the increase of car
traffic, the intensification of social segregation, and also the loss of
physical activity and an increase of obesity \cite{Ewing:2008}. Despite the
importance of the subject, the only studies available at this time are
mostly purely statistical (such as multivariate analysis for example) and we don't have a
clear and simple quantitative model in agreement with data that could
help us in identifying the critical factors of urban sprawl. We won't
discuss all studies about this important problem and we will start
with a simple empirical discussion, followed by scaling arguments. We
will then see how a statistical model can be constructed for
this problem. 

\subsection{Empirical discussion}

\begin{figure}[h!]
\includegraphics[width=0.4\textwidth]{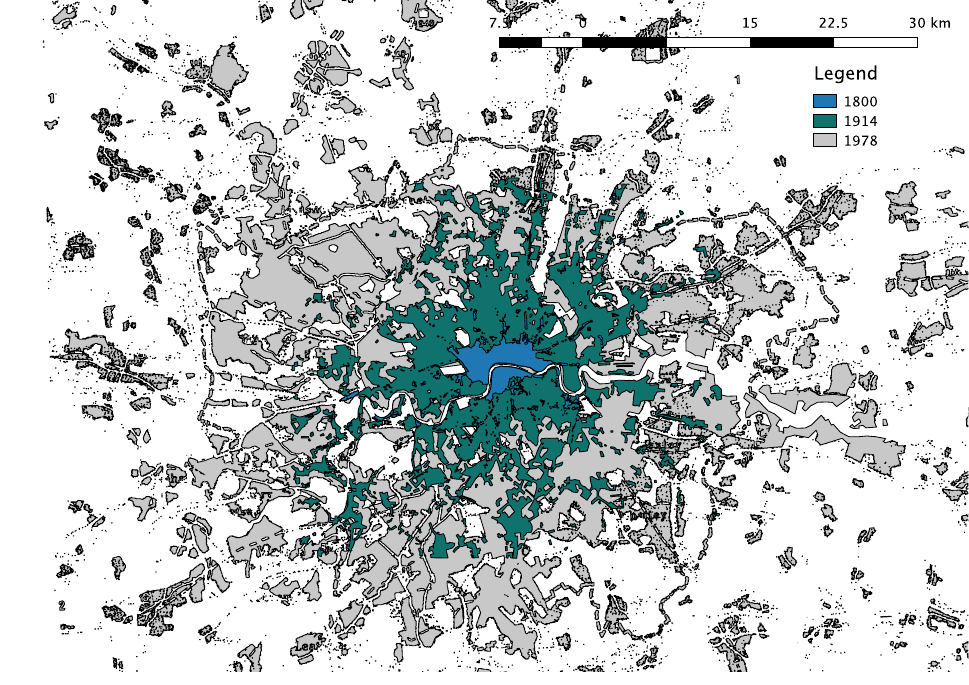}
\caption{ Evolution of the built-up area of London from 1800 to 1978
  (the color-code indicates the age from blue to grey). We observe
  here the growth of a surface but the corresponding equation is not
  known yet. Data from \cite{Angel:book}  and figure courtesy of E. Strano.}
\label{fig:london}
\end{figure}

Visual inspection of the evolution of built-up areas of cities shows the existence of spreading phenomena, and
`hotspots' which suggests that nonlinear mechanisms are present (see
Fig. \ref{fig:london}).
Despite these qualitative discussions, the quantitative
description of this problem is completely open and we could expect
that the physics of growing surface could help. In particular, finding
an equation governing the spatio-temporal evolution of the local
density would be a real breakthrough in our understanding of cities.

\begin{figure}[h!]
\includegraphics[width=0.4\textwidth]{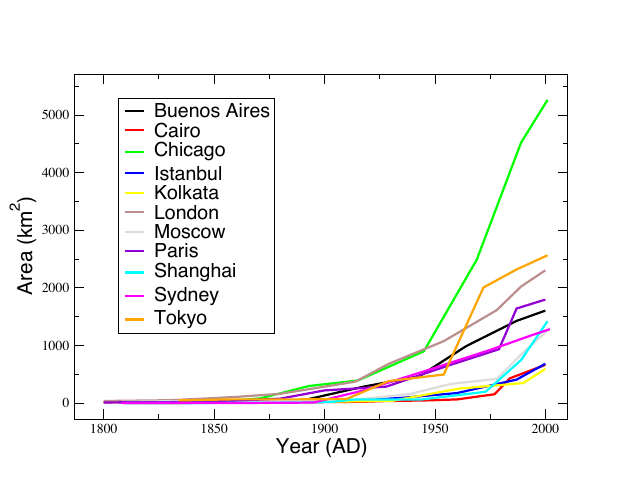}
\caption{ Evolution of the built area of various cities. These curves
  demonstrate an evolution faster than $t$ (typically as $t^2$ for
  most cities). Data from \cite{Angel:book}.}
\label{fig:angel}
\end{figure}

In order to go beyond visual and qualitative explorations, we plot the
built-up area of various cities versus
time and show the results in Fig.~\ref{fig:angel}.
We observe a large variety of behaviors that is difficult to
understand without a theoretical guide, calling for the need of more
empirical analysis and the construction of a theoretical framework. In
particular, the range of variation does not allow here to characterize
precisely the time behavior, even if we can see that in all case the area grows
faster than $t$ (and is consistent in most cases with a $t^2$ behavior) 
implying at least a superdiffusive behavior.

Due to exogenous factors, time is probably not the most relevant variable and it seems
quite natural to study the evolution of the cities surface area versus the population
living in it, a long standing problem in the
field (\cite{Makse:1995}). Here also, we don't have a clear picture yet. For
cities in various countries, we can fit the data with a linear function of the form
\begin{equation}
A=aP
\end{equation}
where $a$ is the inverse of the average density and represents the
typical area per inhabitant. This corresponds to the intuitive
expectation that cities evolve in such a way that their average
population density $\rho = P/A$ remains constant. The quantity $a$ is
the average surface occupied by each individual (the assumption of a
constant density is equivalent to a constant average surface per
capita). We can probably expect this behavior to hold for relatively
small cities where there is an constant increase of the built-up area
when the population is increasing. However, this regime is certainly 
limited as a city cannot extend indefinitely and we expect a later
behavior characterized by a slower increase of the area. In
some other cases, we can
fit the data by a nonlinear function of the form
\begin{equation}
A=aP^\tau
\end{equation}
where $\tau$ is an exponent usually slightly smaller than 1. In the
linear case the density is contant $\rho=P/A=1/a$ and in the nonlinear
case the density increases with population $\rho\sim P^{1-\tau}$:
larger cities are also denser. 

\begin{figure}[ht!]
\includegraphics[width=0.49\linewidth]{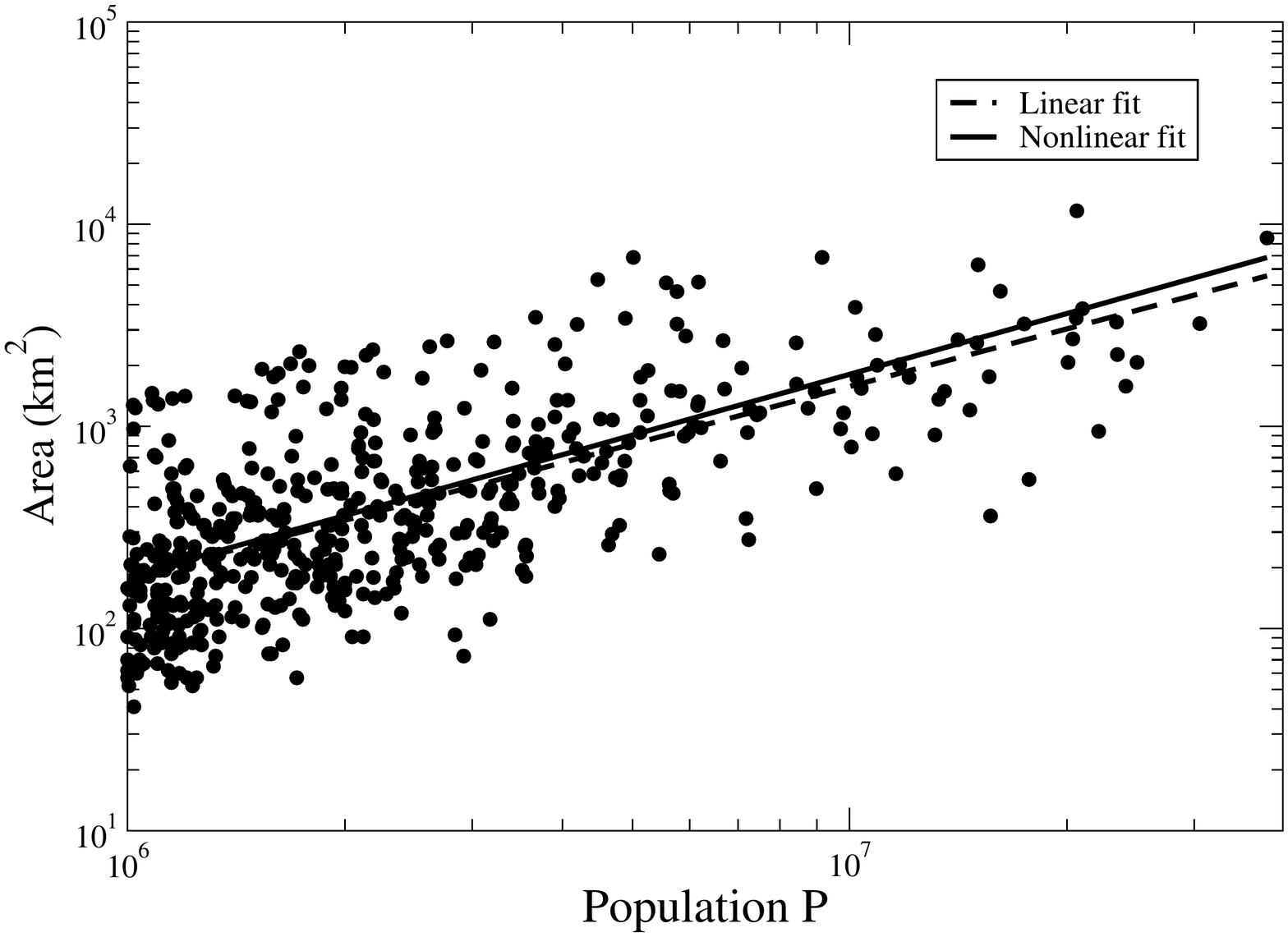}
\includegraphics[width=0.49\linewidth]{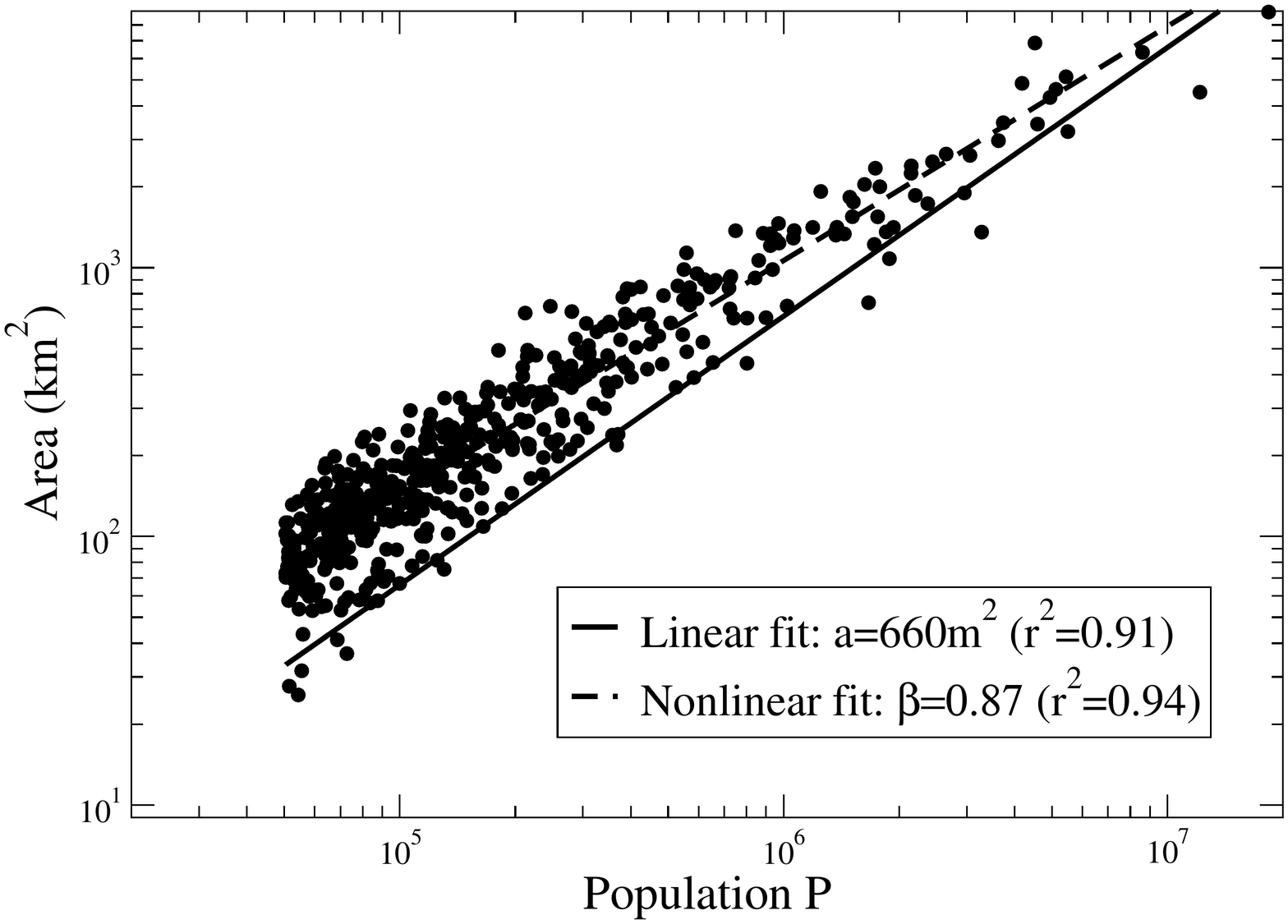}
\caption{Area versus population for different regions in the
  world. (Left) Area of the 494 world cities with population larger than
  1 million inhabitant. Both the linear and nonlinear fit are
  acceptable and very close to each other (data from various sources compiled by Wikipedia). (Right) US
  Urban areas with population over $50,000$ (data from the 2010 Census
  population). Here also both linear and nonlinear fits are acceptable
and a more thorough statistical analysis is needed (such as in
\cite{Leitao:2016} for example).}
\label{fig:density_examples}
\end{figure}

\begin{figure}[ht!]
\includegraphics[width=0.85\linewidth]{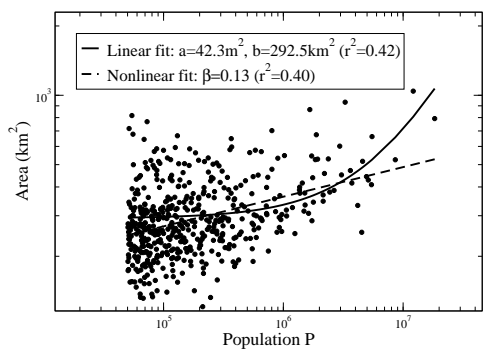}
\caption{Area versus population for urban areas in Japan (data from
  the UN, Demographic Yearbook 2012).}
\label{fig:density_example2}
\end{figure}

Empirically, we do observe that cities
around the world display very different behaviors as shown in Fig.~\ref{fig:density_examples}.
When we mix together different countries of the world by taking urban
areas with population larger than 1million
(Fig.~\ref{fig:density_examples}(left)), the nonlinear fit predicts an
exponent slighly less than one ($\tau\approx 0.95$, $r^2=0.66$) and
with this dataset it seems not possible to conclude between a constant
density or a slightly increasing function of population (with exponent
of order $0.05$). At a smaller scale, in the case of the US for
example (Fig.~\ref{fig:density_examples}(right)), the nonlinear
behavior is also undistinguishable from the linear one. The linear fit
predicts an average area of order $660m^2$ which is large, while the
nonlinear fit gives an exponent $\beta\approx 0.87$.  The linear
fit predicts thus a constant density of order $1,500\mathrm{hab}/km^2$, which is
correct for many cities, but we actually see deviations for large cities
with density much larger for proper cities (and not for urban areas
that mixes heterogeneous zones). For example for NYC, we have a
density of order $10,000\mathrm{hab}/km^2$ or $6,600\mathrm{hab}/km^2$ for San
Francisco. This behavior is however completely different for a country
such as Japan for example (Fig.~\ref{fig:density_example2}) where both
linear and nonlinear fits are not good. The noise in this case cannot
allow to conclude and the data suggests that the area is not a
function of population only.

After this little tour of empirical results, the absence of a clear
stylized fact leaves us in the dark, and theoretical models, even very
simplified ones, could shed some light on this problem. In order to
illustrate how models could help us to understand the data, we will
discuss here two different approaches to this problem: scaling and a
statistical model based on dispersal ideas.

\subsection{Scaling approach}

Bettencourt \cite{Bettencourt:2013} proposed a phenomenological
approach accounting for the value of the exponents governing the
evolution of various quantities versus the population. In this
approach the quantity of interest $Y$ can be any `urban socioeconomic
output' such as income, etc. Empirically we observe in general
\cite{Pumain:2004,Bettencourt:2007} a power law
\begin{align}
Y\sim P^\beta
\end{align}
where $\beta$ is an exponent that can be
different from 1. It was measured in \cite{Bettencourt:2007} for
various quantities and we can distinguish 3 groups according to its
value. For social-related quantities (such as the number of patents, number of
serious crimes, etc.), we observe $\beta>1$ possibly rooted in the fact that interactions in cities grow
very fast with population, typically as $P^2$. In contrast, we also
observe values $\beta<1$, which indicates an economy of scale (road surface,
length of electric cables, etc.), and the last category with exponent
$\beta=1$ comprises essentially quantities that do not depend on the size of the
city (water consumption or other human-dependent quantities for
example).

In order to estimate these exponents (and to obtain as a by-product the behavior
of area with population), Bettencourt assumes that 
the economic output per capita $Y/P$ is proportional to the
  average number of interactions $Y/P\sim g\langle k\rangle$. The
  quantity $g$ is assumed to be constant and the average number of
  interactions is assumed to be $\langle k\rangle\sim\rho\sigma$ where
  $\sigma$ is the `cross-section' of individuals and $\rho=P/A$ the
  average population density. This implies that
\begin{align}
Y/P=G\frac{P}{A}
\end{align}
where $G$ is a constant. In addition, Bettencourt assumes that $Y/P$ is 
of the order of the cost of transport
which is also proportional to the average distance $\ell$ travelled in the
city. The average distance $\ell$ depends on the fractal dimension $H$ of the
  transportation network and Bettencourt writes $\ell\sim A^{H/d}$ (where usually $d=2$ is the
  dimension of the embedding space). We then have 
\begin{align}
Y/P\sim\frac{P}{A}\sim A^{H/d}\\
\Rightarrow A\sim P^{\frac{d}{H+d}}
\end{align}
Usually $d=2$, and in the simplest case of a linear transportation 
network $H=1$, we obtain $\beta=2/3$. This simple approach thus
confirms a sublinear behavior with population but is at this point not
able to explain the large fluctuations and the variety of behavior
observed in real world cities. 


\subsection{A dispersal model}

We are left with the question of constructing a simple model that is
able to explain some of the features observed for the growth of the
surface area of cities. A natural approach is to adapt dispersal models
used in theoretical ecology. These models were developped to describe
the proliferation of animal colonies~\cite{Shigesada:1997, Clark:2001}
and also as simplified models for cancerous tumor growth~\cite{Iwata:2000, Haustein:2012}. The main
feature of dispersal models is the concomitant existence of two growth
mechanisms. The first process is the growth of the main (`primary')
colony, which occurs typically via a reaction-diffusion process (for
example as described by a FKK-like equation
\cite{Fisher:1937,Shigesada:2002}) and leads to a constant growth with
a velocity $c$ that depends on the details of the system. In the case
of urban systems, this process would correspond to new buildings
constructed at the fringe of the city and which are in turn triggering
the construction of new buildings. The second ingredient is {\it
  random dispersal} from the primary colony, which represents the
emergence of secondary settlements. In the urban sprawl case, this
second process corresponds to the creation of small towns in the
periphery of large cities. In real-world systems, dispersion is not
isotropic and is governed mainly by transportation systems (blood
vessels in the case of metastatic tumors, winds and rivers in
ecological examples), but in this first approach we will neglect these
effects. We assume that these secondary colonies grow also at the velocity $c$ and
eventually will coalesce with the primary colony. This image
of a growing city, that eventually coalesces with neighboring small
towns is consistent with our knowledge of urban growth (see for
example the case of West London in \cite{Stanilov:2013}).

We will follow here the approach introduced by
Kawasaki and Shigesada in~\cite{Shigesada:1997, Shigesada:2002} in
order to study this simplified model. We will consider that the primary colony
grows at radial velocity $c$ and  emits  a
secondary colony at a rate $\lambda$ 
 and  at a fixed distance $\ell$ from its border (long-range
dispersal). Besides, we  assume that  each secondary colony  also grows with
the same radial speed $c$ and does not emit tertiary 
 colonies. The dependence of  the emission rate  on the colony size
 is taken into account by the functional form  
\begin{equation}
\lambda(r) = \lambda_0 r^{\theta}~,
\end{equation}
 $r$ being  the radius of the primary colony and 
 $\theta \geq 0$. When $\theta=0$ the growth rate is independent
from the primary colony size,  for $\theta=1$ it is proportional to its
perimeter and for $\theta=2$ to  its area. For cities, we can imagine
that at least the perimeter or the surface are the relevant variables
for triggering new towns in its surrounding and that $\theta\geq 1$ is
therefore probably the most relevant for urban systems.

\begin{figure}[h!]
\begin{center}
\includegraphics[angle=0, width=0.3\textwidth]{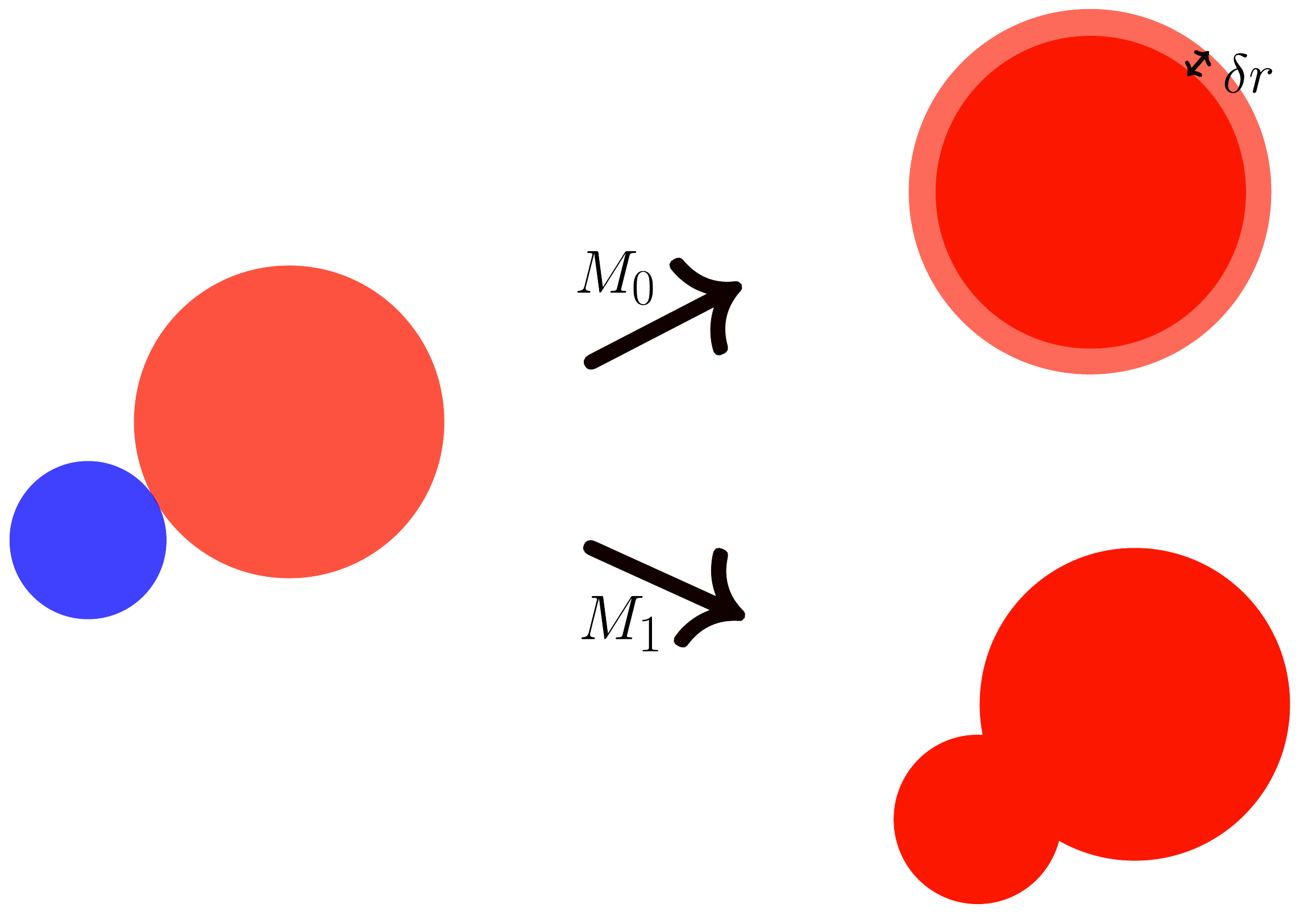} 
\end{center}
\caption{Example of coalescence in  models $M_0$ and $M_1$.
 In $M_0$, the primary colony (in red) remains circular and the area 
 of the secondary colony is  evenly  distributed on the rim; in  $M_1$,
 the shapes are simply  `concatenated'. Figure taken from \cite{Carra:2017}.}
\label{fig:i}
\end{figure}

We consider two variants of this process \cite{Carra:2017} - in a
first version (model $M_0$) we assume that the primary colony remains
circular after the coalescence with a secondary colony. In contrast,
we can consider a modified version of the process (model $M_1$) where 
after the coalescence with a secondary colony, the shape of the  primary colony does not remain circular. This
important difference  is illustrated in the Fig.~\ref{fig:i}.

\begin{figure}[ht!]
\includegraphics[angle=0, width=0.4\textwidth]{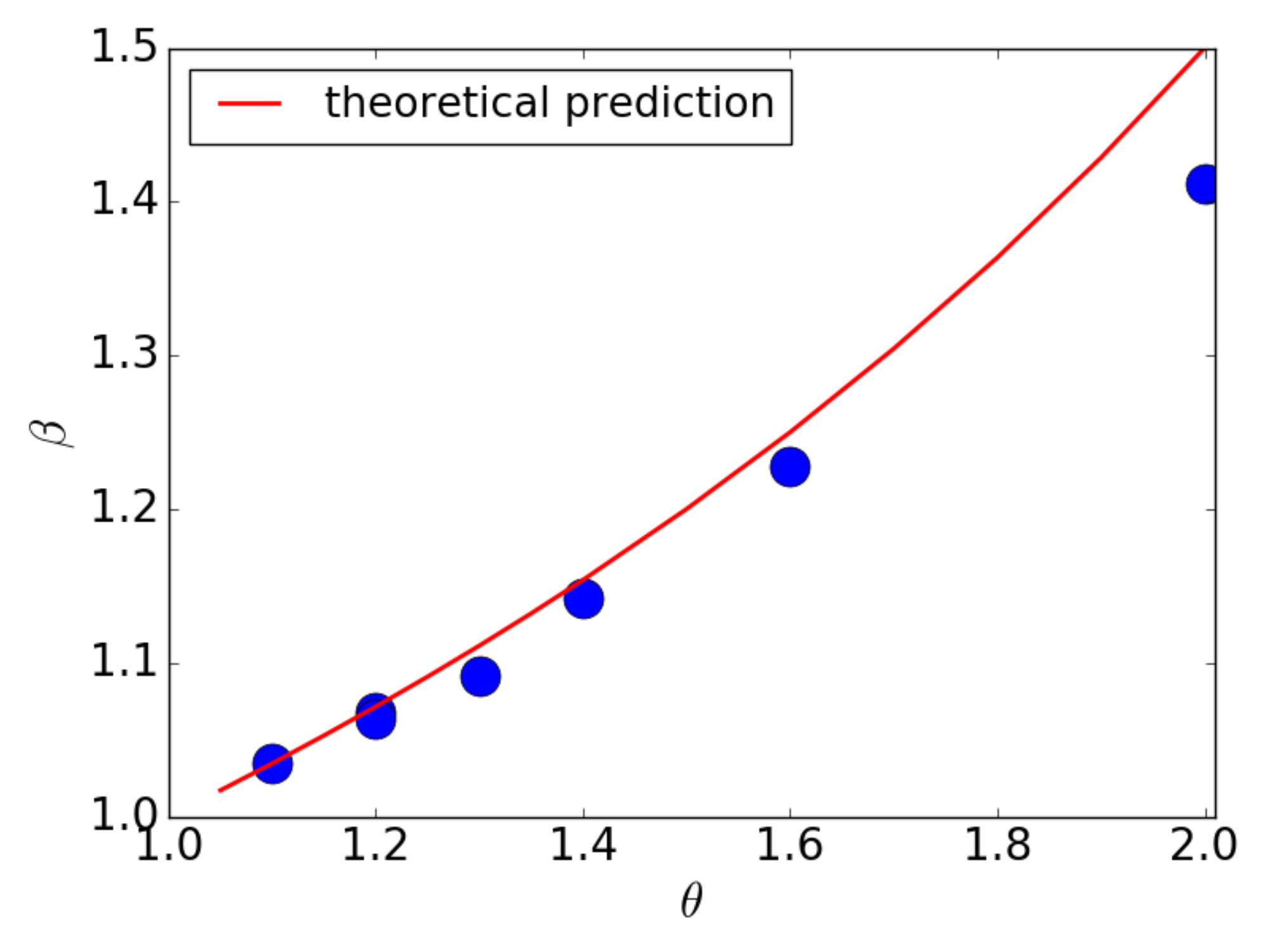} 
\caption{
Plot of  the exponent $\beta$ as a function of  $\theta$, obtained from a power
  law fit on $r(t)$ versus $t$. The theoretical prediction
  Eq.~\eqref{eq:beta} is shown in red. See \cite{Carra:2017} for details.}
\label{fig:3}
\end{figure}

If we denote by $r(t)$ the radius of the primary colony at time $t$
and by $x(t)$  the radius of the colony
absorbed at time $t$, we can show \cite{Carra:2017} that the equations
governing the evolution of these quantities are \cite{Shigesada:2002,Carra:2017}
\begin{align}
\begin{cases}
  \frac{dr}{dt} &= c + \frac{\lambda_0 {r\left(t - 
\frac{x(t)}{c}\right)}^{\theta}}{2\pi r(t)} \left(1 - \frac{\dot{x}(t)}{c}\right)\pi x(t)^2  ~,\\
  \ell &= r(t) - r\left(t - \frac{x(t)}{c}\right) +
  x(t)~.
\end{cases}
\label{eq:Shi_2}
\end{align}
These nonlinear differential equations capture  the physics of the
coalescence and allows us to extract the large time behavior of the
main quantities of interest in this problem. In particular,  assuming 
scaling laws  at  large times, $r(t)  \sim  at^{\beta}$ and $x(t)  \sim
dt^{-\alpha}$,  we obtain 
\begin{equation}
\beta = \frac{3}{4 -\theta}~, \qquad \qquad  \alpha = \beta - 1~.
\label{eq:beta}
\end{equation}
 Note  that for $\theta \rightarrow 4 $, we have $\beta \rightarrow
 \infty$, the radius grows faster than a power law  and 
explodes exponentially. For $\theta = 1$, we obtain
$\alpha =0, \beta =1$ which means that  we have  $x(t) = x^*$
independent of $t$ and a linear behavior of $r(t)$. In this case the
effective radial velocity $c'$ of the primary colony is given by 
\begin{equation}
c' = c + \frac{\lambda_0}{2} {x^*}^2
\end{equation}
and the value of $x^*$ can be obtained by solving 
 Eq.~\eqref{eq:Shi_2} that can be written as
\begin{equation}
\label{eq:c_primo}
\frac{\lambda_0 }{2c} {x^*}^3 + 2{x^*} - \ell = 0~.
\end{equation}
This result,  for the specific case of $\theta = 1$,  was
 first obtained by Shigesada and Kawasaki~\citep{Shigesada:1997}.
More generally, we can test our prediction for $\theta>1$ on numerical simulations,
and in Fig.~\ref{fig:3}, we plot the values of the exponents $\beta$
obtained by power law fits and compare it with the theoretical
prediction Eq.~\eqref{eq:beta}. We observe a good agreement with some deviations
for higher values of $\theta$ which are probably due to finite size
effect.
Even if this model is very simple and is more `metaphorical' than
realistic \cite{Bouchaud:2019}, it suggests that the simple mechanism of
dispersal implies that the growth of the citie's area (for $\theta\geq
1$) is scaling at least as $t^2$ in agreement with the empirical
observations discussed above.

This simple Shigesada-Kawasaki coalescing model is based on the circular
approximation and if we drop this assumption, analytical calculations seem out of
reach. We can however investigate this problem numerically
\cite{Carra:2017} and assume that the area $A$ and the
perimeter $L$ obey to a power law scaling of the form
\begin{equation}
A(t) \sim t^{\mu} \qquad L(t) \sim t^{\nu}~.
\label{eq:beh}
\end{equation}

\begin{figure}[ht!]
\includegraphics[angle=0, width=0.20\textwidth]{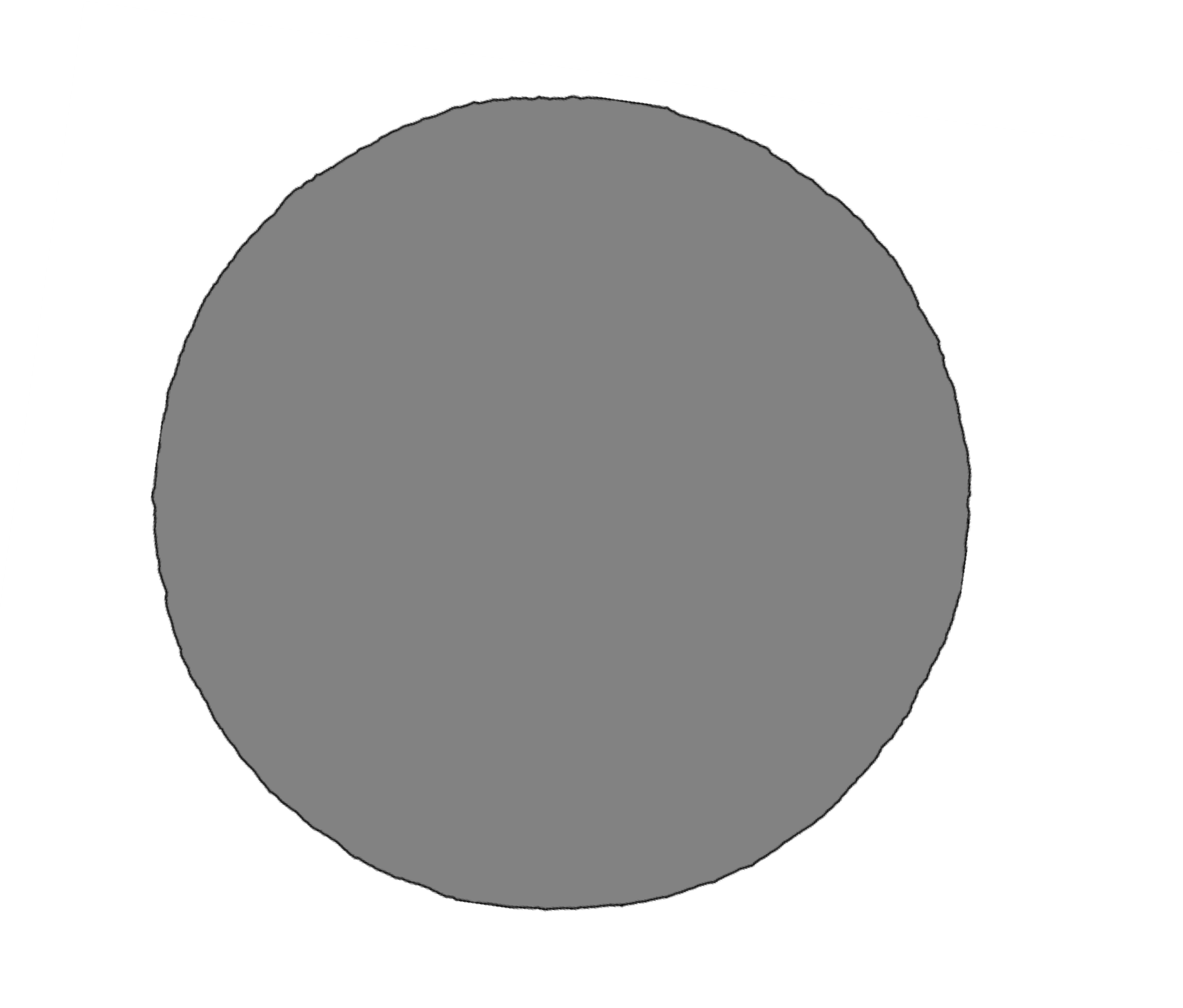}
\includegraphics[angle=0, width=0.20\textwidth]{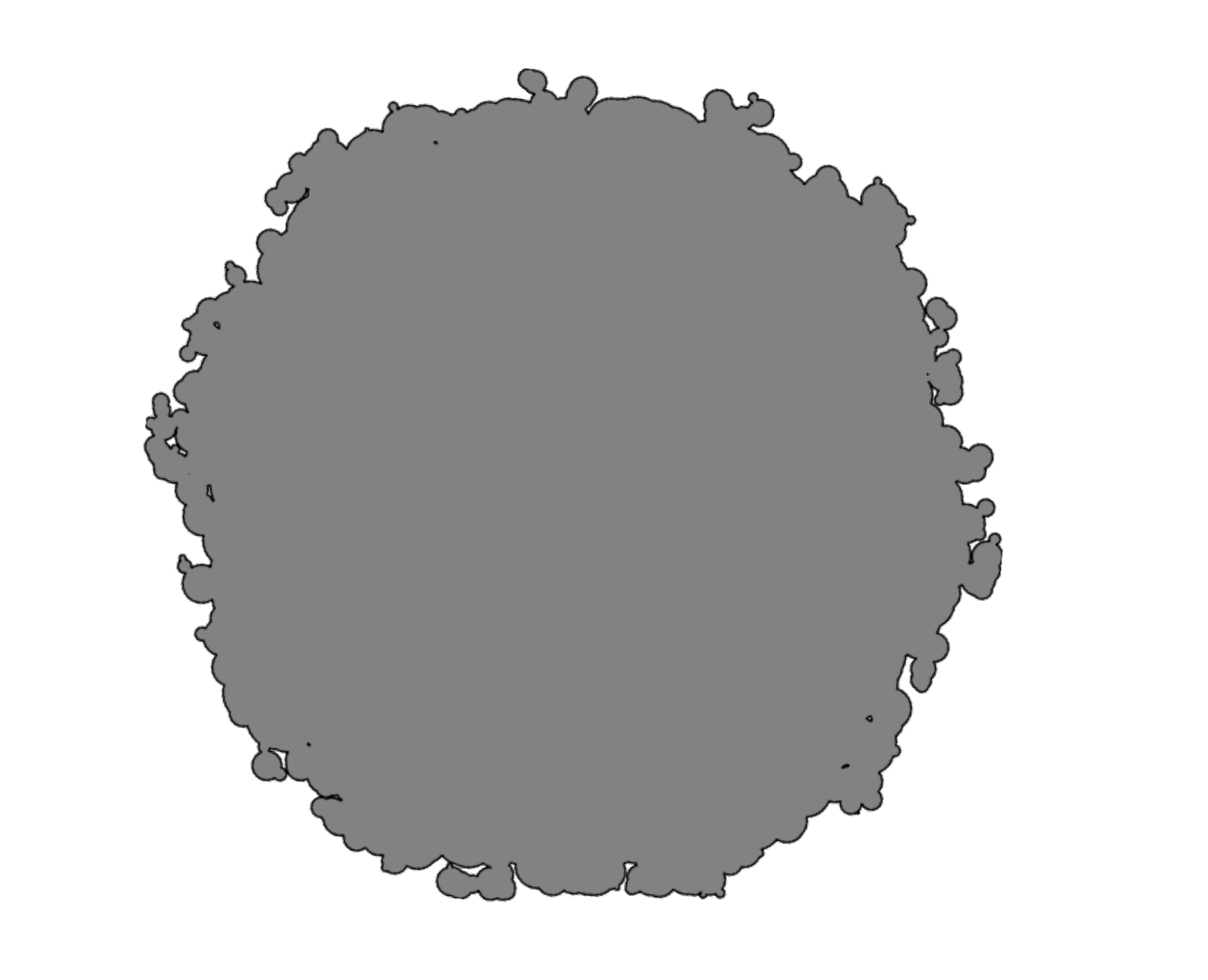} 
\caption{Example of shapes obtained for the model $M_1$. (Left)
  Constant emission rate and $\ell = 10$. (Right) Emission rate
  proportional to the perimeter (Eq.~\ref{eq:perim}), and $\ell = 2$. Figures taken from \cite{Carra:2017}.}
\label{fig:2b5}
\end{figure}

In the case of a constant emission rate $\lambda(r) =\lambda_0$,
numerical results seem to show that $\mu\approx 2$ and $\nu\approx 1$.
The dominant behavior are then $A(t)\sim \pi c^2t^2$ and $L(t) \sim
2\pi ct$. This shows that for $\theta=0$ and for large values of $t$,
the circular approximation is valid: the city grows isotropically and
absorbs neighboring towns. We can also consider the case where the emission rate
$\lambda$ behaves as 
\begin{equation}
\lambda(t) = \lambda_0 L(t)~,
\label{eq:perim}
\end{equation}
where $L(t)$ is the total perimeter of the primary colony at time
$t$. This case corresponds to $\theta=1$ in the model $M_0$.  The
simulations results for the area $A(t)$ and the perimeter $L(t)$ of
the primary colony suggest that we still have $\mu \approx 2$ and
$\nu \approx 1$ as in the $M_0$ model. We can go further and
investigate the prefactors of $A(t)$ and $L(t)$. We recall that for
the $M_0$ model with $\theta = 1$, the radius of the primary colony
increases with an effective radial velocity $c' \neq c $. We can then
study the quantities $A(t)/\pi c'^2 t^2 - 1 $ and
$L(t) / 2 \pi c' t - 1$; if the prefactor is the same of the $M_0$
model we should find (as we did for $\theta = 0$) that these
quantities tend to zero for large values of $t$. In fact we observe
that these two quantities tend to a constant that depends on
$\ell$. These results show that the circular approximation is not
appropriate in the case described by Eq.~\ref{eq:perim} (see Fig.~\ref{fig:2b5}).
When the emission rate is proportional to the perimeter, the circular
approximation thus breaks down and the roughness of the primary colony can
not be discarded, thus modifying the scaling exponents. For cities, we
expect that the rate of creation of new towns depends on the
economical activity of the city and is at least
proportional to its perimeter. These results
show then that -- independently from the anisotropy of transport networks --
cities will in general not grow in an anisotropic way due to the
coalescence with neighboring towns. 

This model is of course very simplified and we are still far from a
quantitative description of cities and their surface area, but we
believe that this type of approach could serve as a basis for more
elaborate ones and eventually to construct the equation governing the
evolution of the area of cities. 

\section{Modeling the population dynamics}

The population of cities varies over several order of magnitudes: from
small towns with hundreds of inhabitants to megacities with more than
10 million inhabitants. This large disparity of
sizes has been noticed for some time already and Zipf uncovered in the
40s the `universal' behavior of the form (see \cite{Zipf:1949})
\begin{align}
P(R)\sim R^{-\nu}
\end{align}             
where $P(R)$ is the population of the city at rank $R$ (cities are
sorted in decreasing order according to their population). Interestingly
enough, this result is robust and valid for different periods in time:
even if there is a non-trivial microdynamics with the rank of cities
changing all the time, Zipf's law remains stable \cite{Batty:2006}. The exponent
$\nu$ is usually close to 1 for most countries which implies that the
population distribution is close to $\rho(P)\sim 1/P^2$ (in general the population size distribution is a
power law with exponent $\kappa=1+1/\nu$). More recent empirical
evidences seem however to show that there are non-negligible fluctuations of
this exponent (see \cite{Soo:2005} for an extensive
study over 73 countries). The Zipf law has a number of
interesting consequences: the ratio of the largest to
second-largest city is given by $P_1/P_2=2^{1/\nu}$; for $\nu<1$ the largest city will have
a population scaling as $P_{max}\sim N^{1/\nu}$;  and the total population of a country
scales as $W\sim N^{1/\nu}$ where $N$ is the number of cities.

Zipf's law for cities triggered a huge number of studies in economics
and also in physics and we discuss here some of the most important
approaches together with an interesting connection with a classical
model of statistical physics. First, Gibrat proposed in 1931 a simple rule stating that
the growth rate of a firm is independent from its size \cite{Gibrat:1931}, and
applied to the growth of city populations, gives the following
equation
\begin{align}
P_i(t+1)=\eta_i(t)P_i(t)
\end{align}
where $P_i(t)$ is the population of city $i$ at time $t$ (usually the year). This equation
basically describes the randomness of births and deaths by introducing
an effective random growth rate $\eta_i(t)$ which is assumed to be independent
from a city to another and without any time correlations. This simple
equation with multiplicative noise naturally leads to a lognormal
population distribution in contrast with empirical results, implying
that the growth of urban areas is not consistent with Gibrat's
law. Several other approaches were then proposed in order to
understand Zipf's law \cite{Marsili:1998,Gabaix:1999}. In particular, 
Gabaix \cite{Gabaix:1999} proposed an alternative approach that is so far considered
as the best explanation for Zipf's law. It is based on the Gibrat
model with the constraint that small cities cannot shrink to
zero. In other words, the process considered by Gabaix is a random
walk with a lower reflecting barrier which can, in some conditions,
produce a power law distribution of populations. In addition to the
reflecting barrier, a necessary condition is a drift towards the
barrier and it is the combination of these two ingredients that can
give rise to a power law \cite{Sornette:1997}. The `regularization' of the
Gibrat model proposed by Gabaix relies on the assumption of a minimum
size of cities and predicts the value $\nu=1$, but cities can
actually disappear and this model is not completely
satisfying. Ideally, we would like an approach based on reasonable
mechanisms, linking micromotives with the large-scale behavior
described by Zipf's law. An interesting step towards such a
description was proposed by Bouchaud and M\'ezard \cite{Bouchaud:2000} (in the context
of the wealth distribution) who discussed the diffusion equation with
noise. This model has implications far beyond cities such as finance,
directed polymers and the KPZ equation, etc. (see \cite{Bouchaud:2000} and references
therein). The diffusion equation with noise for the population
$P_i(t)$ of city $i$ reads in the continuous time limit as
\begin{align}
\frac{dP_i}{dt}=(\eta_i(t)-1)P_i(t)+\sum_jJ_{ij}P_j(t)-J_{ji}P_i(t)
\end{align}
where the first term represents the `internal' growth as given by
Gibrat's law and where the two last terms represent migrations between
cities. We note that in contrast with some other models, the ingredients needed
here are reasonable and correspond to a process that does occur in the
real world. The random variables $\eta_i(t)$ are assumed to be identically
independent Gaussian variables with the same mean and a variance given
by $2\sigma^2$. The flow (per unit time) from city $i$ to $j$ is
denoted by $J_{ji}$ and for a general form of these couplings we are
unable to solve this equation. In the mean-field limit where all
cities are exchanging individuals with each other ($J_{ij}=J/N$ where
$N$ is the number of cities), Bouchaud and M\'ezard could show that the
stationary distribution of the normalized population
$w_i=P_i/\overline{P}$ (where $\overline{P}$ is the average
population) is
\begin{align}
\rho(w)\sim w^{-(1+\kappa)}\mathrm{e}^{-(\kappa-1)/w}
\end{align}
where the exponent is
\begin{align}
\kappa=1+\frac{J}{\sigma^2}
\end{align}
When the migration term $J$ is nonzero, this
regularization changes the lognormal distribution to a power law for
large $w$ and the exponent is between 1 and 2 (for
$J/\sigma^2<1$). The exponent $\kappa$ in this approach is not
universal and depends on the details of the system providing a
possible explanation for the diversity of values observed empirically
\cite{Soo:2005}. This diffusion model suggests an interesting
connection between a central model in statistical physics and the old
problem of the urban population distribution. It also shows that
Zipf's law finds its origin in the interplay between internal random
growth and exchanges between different cities. An important
consequence of this result is that increasing inter-urban mobility should actually
increase $\kappa$ and therefore reduces the heterogeneity of the city
size distribution. Other empirical tests are needed at this point, and
from a theoretical point of view the mean-field assumption is not
obvious. We expect that in general the couplings $J_{ij}$ are not constant and
depend on the distance between cities, which could dramatically alter
the results.

\section{Spatial organization of cities}

In this section, we will focus in the intra-urban scale. In
particular, we will discuss approaches for understanding the spatial
organization of a city. It is a problem of crucial importance as the
location of residences and of the economic activity govern commuting
flows and mobility patterns, a vital ingredient for assessing the
efficiency of infrastructures and for planning. It is thus important to understand how households
and companies choose a certain location and what are the main
driving factors. We will first discuss here the main model used in
urban economics and upon which many variants are constructed. We will
then discuss another approach proposed by Krugman and which is a good
example of a non-equilibrium approach to the city structure. Finally
we will discuss a more recent model proposed in a statistical physics
spirit.

\subsection{The Alonso-Muth-Mills model of urban economics}

The Alonso-Muth-Mills (AMM) model is a pillar of urban economics and constitute
the basis on which most economic models are constructed. In this
respect we believe that it is important to know this model and we
describe its main lines here.  The first
ingredient in this model is a utility function $U$ that describes the
preferences of individuals (or households -- they are usually
considered to be the same in this simple approach) that are considered
to be all equivalent. This function $U$ is
usually assumed to depend on the land consumption $s$ (which
corresponds to the surface area of apartments) and on the composite
commodity $z$ (which corresponds to the money left when rent and
transportation costs are substracted from the income)
\begin{equation}
U=U(z,s)
\end{equation}
This utility has to satisfy the general constraints
\begin{equation}
\frac{\partial U}{\partial s}>0\;\;,\;\;\frac{\partial U}{\partial
  z}>0
\end{equation}
which means that households in general prefer larger apartment
and smaller costs. The budget constraint is given by
\begin{equation}
Y=z+T(x)+R(x)s
\end{equation}
where $Y$ is the income of an household, $T(x)$ the transportation cost to work when
living at location $x$ and $R(x)$ the renting cost per unit area at
$x$. In this simplest version of the AMM model all households are renting their apartments and
all landlords are living out of the city. The problem is then to optimize the utility subject to this budget
constraint
\begin{equation}
\max_{z,s}U(z,s)\;\;\mathrm{subject\;\;to}\;\;Y=z+T(x)+R(x)s
\end{equation}
This is a classical problem which can be solved with Lagrange multipliers, but we 
will here discuss a faster way to obtain general results.  We
introduce the constraint with $z=Y-sR-T$ and we maximize $U(Y-R(x)s-T(x),s)$ with respect to $s$
\begin{equation}
\frac{\partial U}{\partial s}=0=\partial_1 U (-R(x))+\partial_2U
\end{equation}
where $\partial_iU$ denotes the derivative of $U$ with respect to the
$i^{th}$ variable. From this equation, we obtain the renting cost under the form
\begin{equation}
R(x)=\frac{\partial_2U}{\partial_1U}
\label{eq:Rx1}
\end{equation}
An additional requirement is that the maximum utility $U^*$ should
be independent from $x$. If it is not, then
individuals could choose another better location and we wouldn't be at
equilibrium. We thus have to write 
\begin{equation}
\frac{dU^*}{dx}=0=\partial_1U\left(-s\frac{dR}{dx}-R\frac{ds}{dx}-T'(x)\right)+\partial_2U\frac{\partial
  s}{\partial x}
\label{eq:Ux1}
\end{equation}
where the functions $s(x)$ and  $R(x)$ are computed at equilibrium.
Combining Eqs. \eqref{eq:Rx1} and \eqref{eq:Ux1} which are valid
for all $x$, we then obtain the
central result for the AMM model (see for example \cite{Brueckner:1987})
\begin{equation}
\frac{dR}{dx}=-\frac{T'(x)}{s(x)}
\label{eq:amm}
\end{equation}

This relation Eq.~\eqref{eq:amm} allows us to discuss the location of
individuals in the city. For
example, for discussing the impact of income, we assume that
transportation cost are linear in $x$ and then $T'(x)=t$ and that we
have two income categories, rich and poor characterized by their
(fixed) land consumption $s_R$ and $s_P$, and transportation costs
$t_R$ and $t_P$. The category of individuals that lives in a given
area of the city is then the one that is willing to pay more for the
rent at this location. The condition for the poor living in teh center
can be shown to be
\begin{equation}
\frac{t_P}{s_P}>\frac{t_R}{s_R}
\end{equation}
In the opposite case, rich individuals will live in the center as they
are willing to pay more than the poor for this location (see
\cite{Glaeser:2008} for a detailed discussion about this point in the
context of the AMM model).

We refer the interesting reader to the large litterature on the
subject and in particular to the books \cite{Fujita:1989} and
\cite{Fujita:1999}, and we will just make a few remarks about this
approach. First, it assumes that cities are in equilibrium and their
structure optimizes some objective function. Given the large variety of
temporal (and spatial) scales, of processes and
interactions, this assumption seems difficult to accept for cities. Even if we
accept it, we are left with the difficult problem that
some features of the city (such as the population profile for example)
will actually depend on the  precise form of the utility
function. This poses the problem of the choice of the utility function
and how to test it empirically. Also for this model, or for the more
involved Fujita-Ogawa model (see \cite{Fujita:1982} and next
sections), the theoretical predictions are usually not thoroughly tested against data. This is often true
for classical studies of urban economics: theories and assumptions are not
tested against data and serve as conceptual guides to understand some
phenomena but usually with no clear guarantee of their validity.

\subsection{Krugman's model: The Edge-City model}

A non-equilibrium model for the spatial structure of cities and in
particular how the economic activity clusters in specific regions was
proposed by Krugman \cite{Krugman:1996}. This approach is a good
example of a minimal model that could probably be built upon and
amenable to predictions that can be tested. The most important aspect
here is the presence of interactions between firms. These interactions
can lead to a polycentric organization of the city in which businesses
are concentrated in spatially separated clusters. We follow the
discussion proposed by \cite{Krugman:1996} and consider that the city
is one-dimensional and the density of businesses is described by the
function $\rho(x)$ whose integral is assumed to be constant. We assume
that all locations are initially equivalent (which means in particular
that there is relatively uniform transport infrastructure network) and
the attractiveness of a location will depend on the spatial
distribution of businesses.  In order to describe this mathematically,
Krugman introduces a quantity $\Pi(x)$ `a market potential' which describes the level of attractivity of
location $x$ and is given by
\begin{equation}
\Pi(x)=\int K(x-z)\rho(z)\mathrm{d}z
\end{equation}
where the kernel is chosen as $K(x)=A(x)-B(x)$ with functions $A$ and
$B$ that are both decreasing with the
distance. These functions represent the positive and negative
spillovers and how they vary with distance. Businesses will have an incentive to come to  certain
location depending on the level of attractiveness $\Pi$ and the
simplest assumption is to compare $\Pi(x)$ to the average spatial level
$\overline{\Pi}=\int \Pi(x)\rho(x)\mathrm{d}x$ and to write
\begin{equation}
\frac{d\rho(x)}{dt}=\gamma\left[\Pi(x)-\overline{\Pi}\right]
\end{equation}
The density will then increase at locations where
$\Pi(x)>\overline{\Pi}$ and decrease otherwise. This equation
describes in a simple way the evolution of the business density
with time and provide an explanation for the self-organized nature
of cities. Note that since $\Pi(x)$ depends on $\rho$, the quantity  
$\overline{\Pi}=\int \Pi(x)\rho(x)\mathrm{d}x$ is nonlinear in
$\rho$. Numerical simulations indicate that this nonlinear system
indeed leads to various situations with multiple centers at different
locations, depending on the initial conditions. Essentially, if
positive spillovers are larger we will observe bigger clusters of
businesses. We also see that this concentration has a reinforcement
effect: regions with a large market potential will be more attractive
and will therefore grow.

At a more quantitative level, we denote by $\lambda$ the size of spatial
fluctuations and by $r_1, r_2$ the range of positive and negative
spillovers, respectively. For large frequencies, $\lambda\ll r_1,r_2$
there is basically a compensation effect of positive and negative
spillover and we do not expect the growth rate to be large. Also, for
very low frequency fluctuations such as for example around a maximum of $\rho(x)$ which
decays slowly around $x$, and for strong enough negative spillover,
the growth rate at $x$ will be negative. Both low and high frequencies
have thus negligible growth rate and it is natural to expect a frequency
with a maximal growth rate, well-tuned to the spatial decay of positive
and negative spillovers, leading to a specific spatial pattern in
this city. In order to get a quick analytical insight, we can linearize the equation for $\rho(x)$
around the flat city $\rho(x)=\rho_0+\delta\rho(x)$ and find
\begin{equation}
\frac{d\delta\rho}{dt}\simeq \gamma\int
K(x-z)\delta\rho(z)\mathrm{d}z+{\cal O}(\delta\rho^2)
\end{equation}
where we choose the normalization $\int\rho(x)=1$. Using the Fourier
transform
\begin{equation}
\delta\rho(k)=\int \mathrm{e}^{ikx}\delta\rho(x)\mathrm{d}x
\end{equation}
we obtain by integrating the linear differential equation
\begin{equation}
\delta\rho(k)\sim \mathrm{e}^{\gamma\hat{K}(k)}
\end{equation}
where $\hat{K}(k)$ is the Fourier transform of the kernel $K(x)$. This
expression shows that the Fourier mode $k^*$ for which $\hat{K}(k)$ is
maximum will develop faster and will lead to the appearance of a
spatial pattern characterized by $k^*$. We can obtain explicit
expressions with the choice
$K(x)=A\mathrm{e}^{-|x|/r_1}-B\mathrm{e}^{-|x|/r_2}$ 
which leads to
\begin{equation}
\frac{d\delta\rho(k,t)}{dt}=2\gamma\left[\frac{Ar_1}{1+(r_1k)^2}-\frac{Br_2}{1+(r_2k)^2}\right]\delta\rho(k,t)
\end{equation}
We thus have for each mode a growth rate proportional to 
\begin{equation}
\Lambda(k)=\frac{Ar_1}{1+(r_1k)^2}-\frac{Br_2}{1+(r_2k)^2}
\end{equation}
which satisfies $\Lambda(0)=A-B$ which is negative for large enough
$B$ and $\Lambda (k)\to 0$ for $k$ large. We thus expect in general a
maximum $\Lambda^*$ for a value $k^*$ that can be easily
computed. This value $k^*$ depends here on $r_1,r_2,A$ and $B$ and is
thus finite and independent from the city size. For a one dimensional
city of size $L$, the number of business clsuters is then given by
\begin{equation}
H\sim Lk^*
\end{equation}
This simple model thus predicts a linear increase of the number of activity centers
(or `hostspots') with city size, but doesn't explain in particular how
this quantity scales with population. Indeed, new datasources such as cell phone networks, employment data, smart
card transactions and taxi GPS trajectories provide a lot of
information about cities and their structure. In particular, it has
been shown \cite{Louf:2013,Louail:2014} that the number $H$ of activity
centers (or `hotspots') is scaling with population as 
\begin{align}
H\sim P^\sigma
\label{eq:h}
\end{align}
where the exponent $\sigma$ is found to be around $0.5-0.6$. The
number of these hotspots thus scales sublinearly with the population
size, a result that will serve as a guide for constructing a
theoretical model. We thus see that Krugman's model is not able to
explain this behavior. Although it is interesting to
see how simple nonlinear effects give rise to a nontrivial spatial
pattern, this model doesn't produce at this stage predictions that are
directly testable on empirical data.

\subsection{A variant of Fujita-Ogawa}

\begin{figure}[h!]
\includegraphics[width=0.4\textwidth]{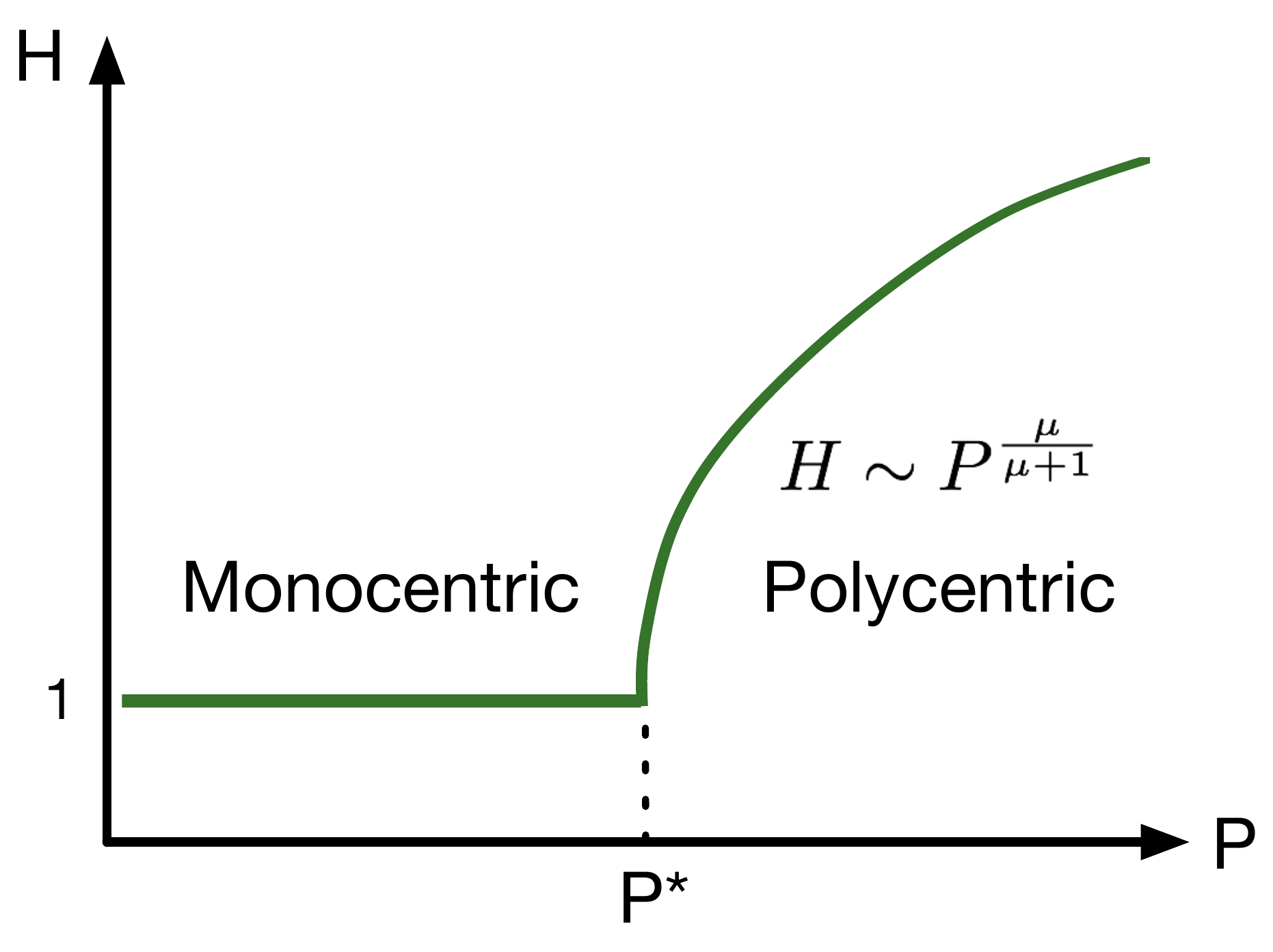}
\caption{Number of activity centers versus the population for the
model discussed in the text: we observe a monocentric-polycentric
transition for a value $P^*$ that depends on the city.}
\label{fig:poly}
\end{figure}

The Krugman model discussed above is unable to explain the empirical
behavior Eq.~\eqref{eq:h}. We will show here how a simplified variant of the
Fujita-Ogawa model \cite{Fujita:1982} can actually help us to understand this empirical result. 
We start from the standard economical assumption that an 
invidivual will choose a residence located
at $x$ and to work at location $y$ such that the quantity
\begin{align}
Z(x,y)=W(y)-C_R(x)-T(x,y)
\end{align}
is maximum. The quantity $W(y)$ is the typical wage earned at location
$y$, $C_R(x)$ is the rent cost at $x$, and $T(x,y)$ is the
transportation cost to go from $x$ to $y$ and is usually taken to be
proportional to the time $\tau(x,y)$ spent to cover this distance. In
their original paper, Fujita and Ogawa \cite{Fujita:1982} couldn't
find the general solution, but tested the stability of some specific
urban forms. For example, by neglecting congestion and writing
$\tau(x,y)=d(x,y)/\overline{v}$ (where $d(x,y)$ is the euclidean
distance between $x$ and $y$ and $\overline{v}$ is the free flow
average velocity), they could show that if the transportation cost for
the typical interaction distance between companies becomes too large,
the monocentric organization with a central business district
surrounded by residential areas is unstable. However, this formalism
does not allow to predict the resulting urban structure, and we have to
simplify it in order to reach testable predictions. First, we assume
that each agent has a residence located at random. Second, a quantity
as complex as the wage results from a large number of interactions and
factors, and it is tempting -- in the spirit of random Hamiltonians
for heavy ions \cite{Dyson:1962} -- to replace this complex quantity
by a random number $\phi(y)$: $W(y)=s\phi(y)$ where $s$ sets the
salary scale (and with a certain distribution of $\phi$ which is
irrelevant at this point). Last but not least, we assume that most of
the displacements are made by car and we include congestion effect
which implies that the time $\tau(x,y)$ depends on the traffic
$Q(x,y)$ as described by the simple Bureau of Public Roads function
(see for example \cite{Branston:1976})
\begin{align}
\tau(x,y)=\frac{d(x,y)}{\overline{v}}\left[1+\left(\frac{Q(x,y)}{C}\right)^\mu\right]
\end{align}
where $C$ is the road system capacity, and $\mu$ an exponent, usually
between 2 and 5. These ingredients put together allow for a simple mean-field analysis showing that the
monocentric organization is unstable for a value of the population
larger than a threshold $P^*$ (which depends on the details of the city) and
that the number $H$ of distinct activity centers is given by 
\begin{align}
H\sim\left(\frac{P}{P^*}\right)^{\frac{\mu}{\mu+1}}
\end{align}
(see Fig.~\ref{fig:poly}). This simplified model thus predicts a sublinear
behavior for the number of activity centers with an exponent given by
$\sigma=\mu/(\mu+1)$ (for $\mu\approx 2$, we then recover the
empirical value $\sigma\approx 0.6$, and for $\mu=0$ we have
$H\sim{\cal O}(1)$ in
the absence of congestion). We observe that
whatever the value of $\mu$, the behavior is sublinear, and congestion
appears as a critical factor that shapes the structure of the city and
favors the appearance of new activity centers.

Finally, we note that knowing both the residence and workplace
locations allows to discuss multiple aspects of the commuting to work
\cite{Louf:2014}. In particular, we can estimate the total commuting
time and the quantity of CO$_2$ emitted by cars (see also \cite{Verbavatz:2019}).

\section{Averaging over many cities ?}

Various studies of cities such as scaling
\cite{Bettencourt:2007} use data from different cities at different
times and plot some quantity versus population. For many quantities
(denoted $Y$ here), we observe a power law of the form $Y\sim P^\beta$
where $P$ is the population of the city \cite{Bettencourt:2007}. As we
already discussed above, we can distinguish different behavior
according to the value of $\beta$, in particular if it is larger than
one or not. Once we have measured this scaling form, we could in principle
use it for predicting the behavior of an individual city
when its population changes. In order to illustrate the possible
problems of such an approach we consider the particular case of 
delays due to traffic congestion and analyze a dataset for 101 US
cities in the time range 1982-2014 \cite{Depersin:2018}. This is a
particularly interesting dataset as it is both transversal -- it
contains many cities -- and longitudinal -- for each city we have the
temporal evolution of the delay.

The scaling form obtained by agglomerating all the available data for
different cities and for different years displays a nonlinear
behavior, seemingly in agreement with general empirical results about
scaling \cite{Bettencourt:2007}.
\begin{figure}[!h]
\includegraphics[scale=0.4]{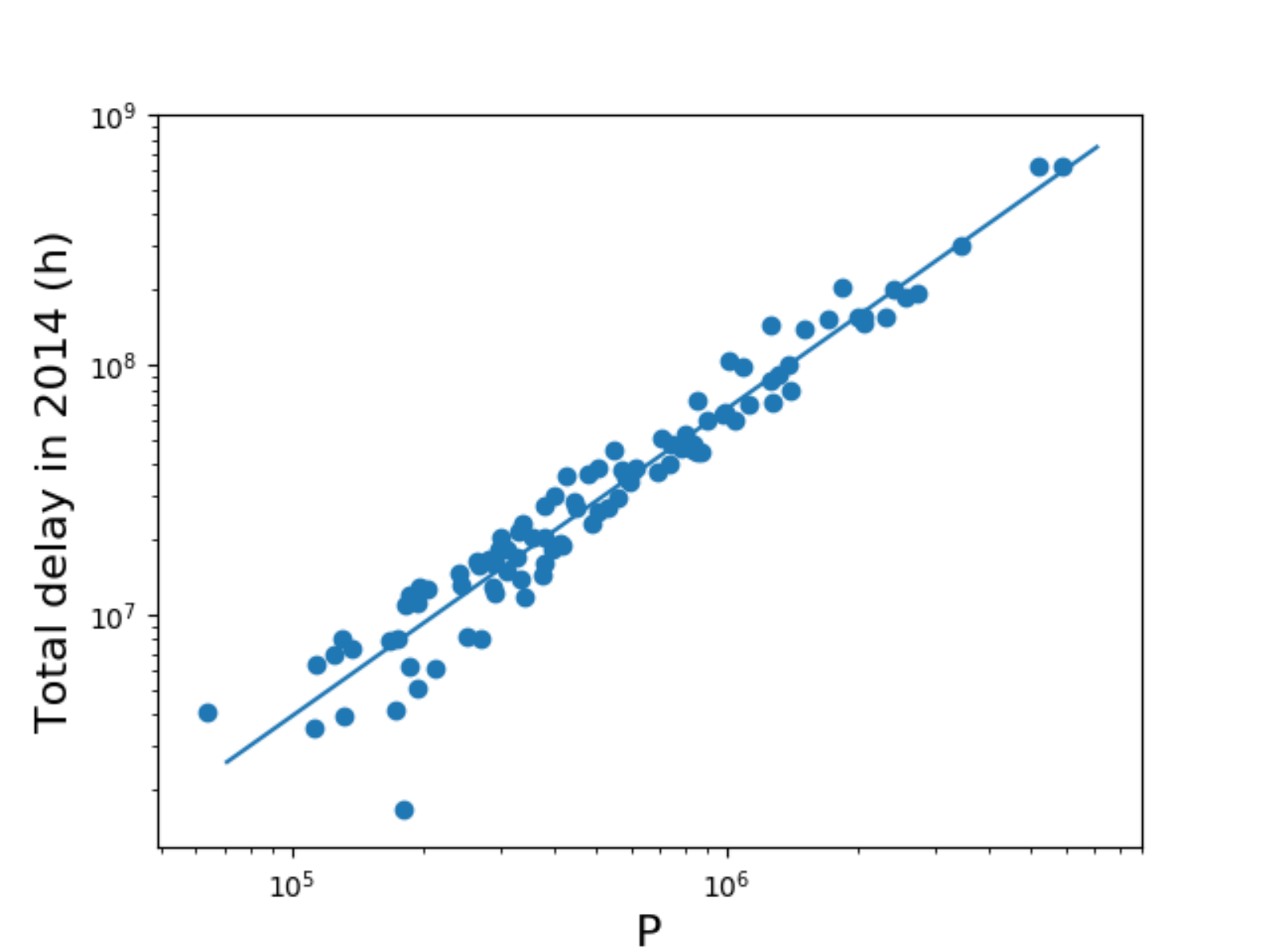}
\includegraphics[scale=0.4]{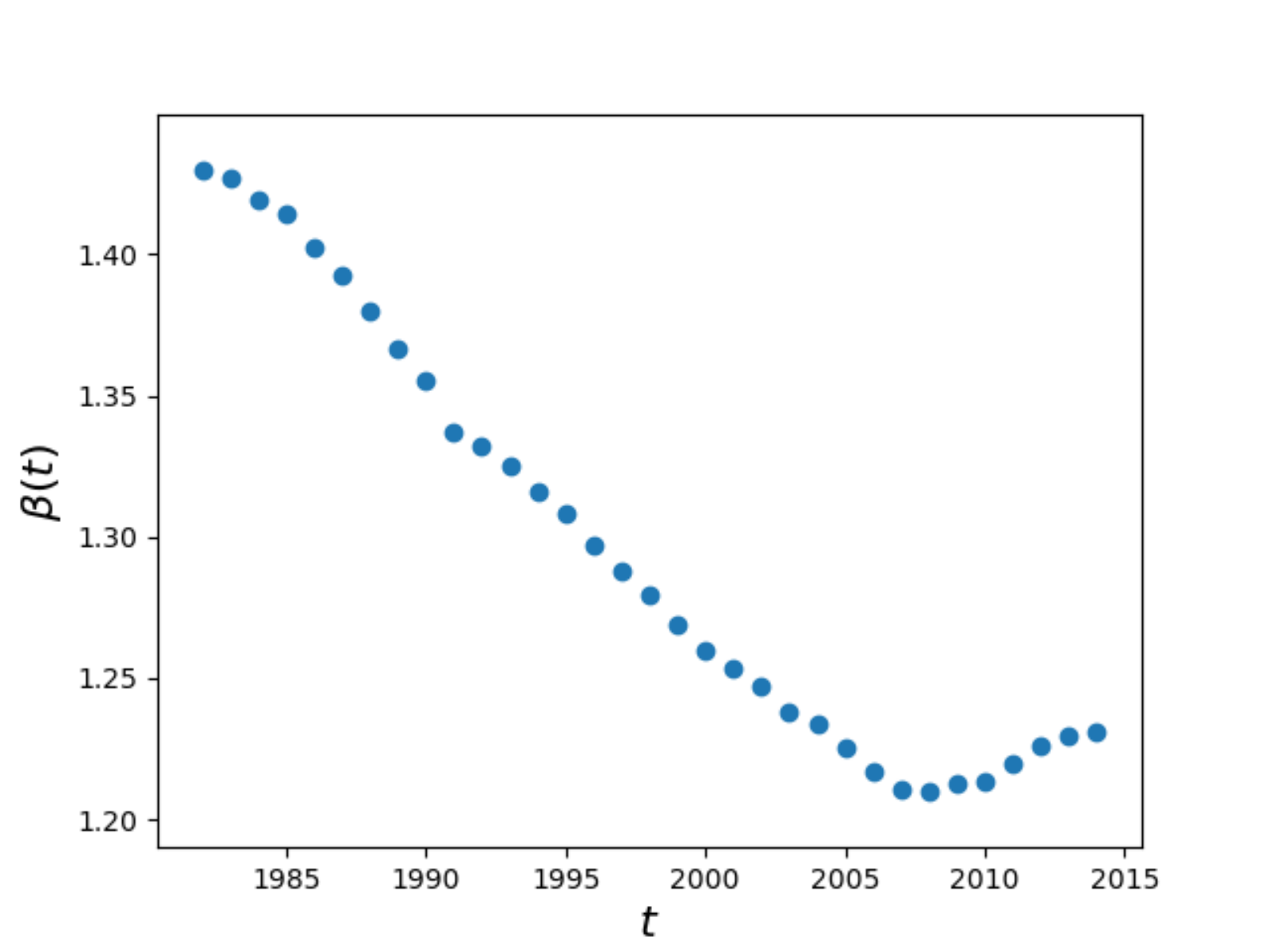}
\caption{(Top) Plot of the annual delay $\delta\tau$ versus the number
  of drivers $P$ for all cities in 2014. The straight line is
  a power law fit in this loglog representation and gives an exponent
  value $\beta\approx 1.23$ (and $R^2=0.93$). (Bottom) Scaling
  exponent $\beta(t)$ for the delay computed for each year separately,
  from 1982 to 2014. All these values are consistent with a
  superlinear behavior found in \cite{Chang:2017}. Figure taken from \cite{Depersin:2018}.}
\label{fig:2014}
\end{figure}
More precisely, we can first compute the scaling exponent by mixing together all
cities but for a given year. We can also compute
this exponent for each year and we observe an exponent whose value varies in the range
$[1.2,1.4]$ for years from 1982 to 2014 (see
Fig.~\ref{fig:2014}). Finally, we can consider all cities and for all years and we then obain a delay in traffic jams scaling as
\begin{align}
\delta\tau\sim P^\beta
\end{align}
with $\beta\approx 1.36$ consistent with a superlinear behavior. 

\begin{figure}[ht!]
\includegraphics[width=1.0\columnwidth]{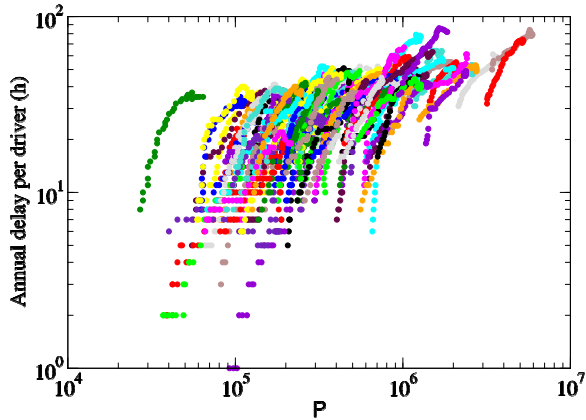}
\caption{Annual delay per capita $\delta\tau/P$ versus $P$ for all the
  101 cities and for all years (1982 -- 2014). The points are colored according to the
city they describe (one color per city). As we discuss in the text there is no
obvious relation between the global power-law scaling and the individual
behavior of cities. Figure taken from \cite{Depersin:2018}.}
\label{fig:deper}
\end{figure}

However, if we don't average over the different cities, we observe the
behaviors shown by different colors in the Fig.~\ref{fig:deper}.
The different cities display then a variety of behaviors ranging from
pure power laws with exponents larger than one, or two power laws,
etc. (see \cite{Depersin:2018} for details). The scaling obtained by
averaging over all cities appears then completely unrelated to the
dynamics of individual cities when their population grow. There seems
to be no simple scaling at the individual city level but a variety of
behaviors. In the language of statistical physics, the delay is not a
state function determined by the population only, and displays some
sort of aging effect where it depends not on the population only but
also on time, and probably on the whole history of the city. This idea
of path-dependency is natural for many complex systems, and in
particular in statistical physics (such as spin-glasses
\cite{Bouchaud:1998} for example). The delay is not a simple function
of population as it is usually assumed for the scaling approach for
cities, and if we reflect upon this idea, it doesn’t make sense in
general to compare two cities having the same population but at very
different dates: both central Paris and Phoenix (AZ) had a population
of about 1 million inhabitants, the former in 1840 and the latter in
1990. It is very likely that the dynamics – for many quantities – from
1840 in Paris will be very different from the one starting in 1990 in
Phoenix, implying that the usual scaling form does not apply in
general. This discussion on congestion induced delays highlights the
risk of agglomerating data for different cities and to consider that
cities are scaled-up versions of each other: there are strong
constraints for being allowed to do that such as path-independence,
which is apparently not satisfied in the case of congestion and which
should be checked in each case. Beyond scaling, these results also
pose the challenging problem of using transversal data (ie. for
different cities) in order to get some information about the temporal
series for individual cities. This is a fundamental problem that needs
to be clarified when looking for generic properties of cities.

\section{Perspectives}

Many aspects and studies about cities were not addressed in this
paper: the evolution of infrastructure networks
\cite{Strano:2012,Levinson:2006}, the coupling between networks
\cite{Kivela:2014}, multimodality \cite{Gallotti:2014,Strano:2015},
heat islands and urban forms \cite{Sobstyl:2018}, etc. An important
point here was to show through various examples how a combination of
empirical results, economical ingredients, and statistical physics
tools can lead to parsimonious models with predictions in agreement
with observations. In particular, describing individual actions by
stochastic processes and replacing complex quantities resulting from
the interactions of several agents by random variables, seem to be a
good way to construct models for understanding the evolution of cities
and in agreement with empirical observations.

We however still have to think about the best approach for
understanding complex systems such as cities. Is it possible to
construct a generic model of cities that can predict general trends
and behaviors ? In this perspective, a particular city would then just
be described by this generic model subject to local spatial and
historical constraints. In other words, this assumption means that all
cities belong to the same species but each individual city evolved on a
different substrate. As we saw in this short review, this possibility
and the existence of generic behaviors is not empirically clear. We do
observe generic properties such as the Zipf law and mixing together
different cities doesn't seem to pose a problem in this case, in
contrast with the exemple of congestion induced delays. There is
however the risk that this debate becomes quickly outdated as machine
learning displays impressive results in practical applications. At
least, we could always hope that parsimonious models have the
advantage to provide a simple language for making sense of the vast
amount of data and to identify critical factors for the evolution of
these systems.


\bibliographystyle{unsrtnat} 
\bibliography{bibfile}		         

\begin{thebibliography}{64}
\providecommand{\natexlab}[1]{#1}
\providecommand{\url}[1]{\texttt{#1}}
\expandafter\ifx\csname urlstyle\endcsname\relax
  \providecommand{\doi}[1]{doi: #1}\else
  \providecommand{\doi}{doi: \begingroup \urlstyle{rm}\Url}\fi

\bibitem[UN:()]{UN:2018}
United nations, world urbanization prospects, 2018.
\newblock URL \url{https://esa.un.org/unpd/wup/}.

\bibitem[Barthelemy(2016)]{Barthelemy:2016}
Marc Barthelemy.
\newblock \emph{The structure and dynamics of cities}.
\newblock Cambridge University Press, 2016.

\bibitem[Fujita(1989)]{Fujita:1989}
Masahisa Fujita.
\newblock \emph{Urban economic theory: land use and city size}.
\newblock Cambridge University Press, 1989.

\bibitem[Von~Thunen and Hall(1966)]{Vonthunen:1966}
Johann~Heinrich Von~Thunen and Peter~Geoffrey Hall.
\newblock \emph{Isolated state}.
\newblock Pergamon, 1966.

\bibitem[Fujita and Ogawa(1982)]{Fujita:1982}
Masahisa Fujita and Hideaki Ogawa.
\newblock Multiple equilibria and structural transition of non-monocentric
  urban configurations.
\newblock \emph{Regional science and urban economics}, 12\penalty0
  (2):\penalty0 161--196, 1982.

\bibitem[Fujita et~al.(2001)Fujita, Krugman, and Venables]{Fujita:1999}
Masahisa Fujita, Paul~R Krugman, and Anthony~J Venables.
\newblock \emph{The spatial economy: Cities, regions, and international trade}.
\newblock MIT press, 2001.

\bibitem[Batty(2008)]{Batty:2008c}
Michael Batty.
\newblock Fifty years of urban modeling: Macro-statics to micro-dynamics.
\newblock In \emph{The dynamics of complex urban systems}, pages 1--20.
  Springer, 2008.

\bibitem[Pumain and Sanders(2013)]{Pumain:2013}
Denise Pumain and Lena Sanders.
\newblock Theoretical principles in interurban simulation models: a comparison.
\newblock \emph{Environment and Planning A}, 45\penalty0 (9):\penalty0
  2243--2260, 2013.

\bibitem[Batty and Longley(1994)]{Batty:1994}
Michael Batty and Paul~A Longley.
\newblock \emph{Fractal cities: a geometry of form and function}.
\newblock Academic Press, 1994.

\bibitem[Tannier and Pumain(2005)]{Tannier:2005}
C{\'e}cile Tannier and Denise Pumain.
\newblock Fractals in urban geography: a theoretical outline and an empirical
  example.
\newblock \emph{Cybergeo: European Journal of Geography}, 2005.

\bibitem[Witten and Sander(1983)]{Witten:1983}
Thomas~A Witten and Leonard~M Sander.
\newblock Diffusion-limited aggregation.
\newblock \emph{Physical Review B}, 27\penalty0 (9):\penalty0 5686, 1983.

\bibitem[Makse et~al.(1995)Makse, Havlin, and Stanley]{Makse:1995}
Hern{\'e}n~A Makse, Shlomo Havlin, and HE~Stanley.
\newblock Modelling urban growth.
\newblock \emph{Nature}, 377:\penalty0 19, 1995.

\bibitem[Makse et~al.(1998)Makse, Andrade, Batty, Havlin, Stanley,
  et~al.]{Makse:1998}
Hern{\'a}n~A Makse, Jos{\'e}~S Andrade, Michael Batty, Shlomo Havlin, H~Eugene
  Stanley, et~al.
\newblock Modeling urban growth patterns with correlated percolation.
\newblock \emph{Physical Review E}, 58\penalty0 (6):\penalty0 7054, 1998.

\bibitem[Rozenfeld et~al.(2008)Rozenfeld, Rybski, Andrade, Batty, Stanley, and
  Makse]{Rozenfeld:2008}
Hern{\'a}n~D Rozenfeld, Diego Rybski, Jos{\'e}~S Andrade, Michael Batty,
  H~Eugene Stanley, and Hern{\'a}n~A Makse.
\newblock Laws of population growth.
\newblock \emph{Proceedings of the National Academy of Sciences}, 105\penalty0
  (48):\penalty0 18702--18707, 2008.

\bibitem[Schelling(1971)]{Schelling:1971}
Thomas~C Schelling.
\newblock Dynamic models of segregation†.
\newblock \emph{Journal of mathematical sociology}, 1\penalty0 (2):\penalty0
  143--186, 1971.

\bibitem[Vinkovi{\'c} and Kirman(2006)]{Vinkovic:2006}
Dejan Vinkovi{\'c} and Alan Kirman.
\newblock A physical analogue of the schelling model.
\newblock \emph{Proceedings of the National Academy of Sciences}, 103\penalty0
  (51):\penalty0 19261--19265, 2006.

\bibitem[Grauwin et~al.(2009)Grauwin, Bertin, Lemoy, and Jensen]{Grauwin:2009}
S{\'e}bastian Grauwin, Eric Bertin, R{\'e}mi Lemoy, and Pablo Jensen.
\newblock Competition between collective and individual dynamics.
\newblock \emph{Proceedings of the National Academy of Sciences}, 106\penalty0
  (49):\penalty0 20622--20626, 2009.

\bibitem[Gauvin et~al.(2009)Gauvin, Vannimenus, and Nadal]{Gauvin:2009}
Laetitia Gauvin, Jean Vannimenus, and J-P Nadal.
\newblock Phase diagram of a schelling segregation model.
\newblock \emph{The European Physical Journal B}, 70\penalty0 (2):\penalty0
  293--304, 2009.

\bibitem[Dall’Asta et~al.(2008)Dall’Asta, Castellano, and
  Marsili]{Dallasta:2008}
Luca Dall’Asta, Claudio Castellano, and Matteo Marsili.
\newblock Statistical physics of the schelling model of segregation.
\newblock \emph{Journal of Statistical Mechanics: Theory and Experiment},
  2008\penalty0 (07):\penalty0 L07002, 2008.

\bibitem[Jensen et~al.(2018)Jensen, Matreux, Cambe, Larralde, and
  Bertin]{Jensen:2018}
Pablo Jensen, Thomas Matreux, Jordan Cambe, Hernan Larralde, and Eric Bertin.
\newblock Giant catalytic effect of altruists in schelling’s segregation
  model.
\newblock \emph{Physical Review Letters}, 120\penalty0 (20):\penalty0 208301,
  2018.

\bibitem[Batty(2013)]{Batty:2013}
Michael Batty.
\newblock \emph{The New Science of Cities}.
\newblock MIT Press, 2013.

\bibitem[Bettencourt et~al.(2013)Bettencourt, Lobo, and
  Youn]{Bettencourt:2013b}
Luis Bettencourt, Jose Lobo, and Hyejin Youn.
\newblock The hypothesis of urban scaling: formalization, implications and
  challenges.
\newblock \emph{arXiv preprint arXiv:1301.5919}, 2013.

\bibitem[Brueckner et~al.(2000)]{Brueckner:2000}
Jan~K Brueckner et~al.
\newblock Urban sprawl: diagnosis and remedies.
\newblock \emph{International regional science review}, 23\penalty0
  (2):\penalty0 160--171, 2000.

\bibitem[Ewing et~al.(2008)Ewing, Schmid, Killingsworth, Zlot, and
  Raudenbush]{Ewing:2008}
Reid Ewing, Tom Schmid, Richard Killingsworth, Amy Zlot, and Stephen
  Raudenbush.
\newblock Relationship between urban sprawl and physical activity, obesity, and
  morbidity.
\newblock In \emph{Urban Ecology}, pages 567--582. Springer, 2008.

\bibitem[Angel et~al.(2005)Angel, Sheppard, Civco, Buckley, Chabaeva, Gitlin,
  Kraley, Parent, and Perlin]{Angel:book}
Shlomo Angel, Stephen Sheppard, Daniel~L Civco, Robert Buckley, Anna Chabaeva,
  Lucy Gitlin, Alison Kraley, Jason Parent, and Micah Perlin.
\newblock \emph{The dynamics of global urban expansion}.
\newblock Citeseer, 2005.

\bibitem[Leit{\~a}o et~al.(2016)Leit{\~a}o, Miotto, Gerlach, and
  Altmann]{Leitao:2016}
Jorge~C Leit{\~a}o, Jos{\'e}~Mar{\'\i}a Miotto, Martin Gerlach, and Eduardo~G
  Altmann.
\newblock Is this scaling nonlinear?
\newblock \emph{Royal Society open science}, 3\penalty0 (7):\penalty0 150649,
  2016.

\bibitem[Bettencourt(2013)]{Bettencourt:2013}
Lu{\'\i}s~MA Bettencourt.
\newblock The origins of scaling in cities.
\newblock \emph{science}, 340\penalty0 (6139):\penalty0 1438--1441, 2013.

\bibitem[Pumain(2004)]{Pumain:2004}
Denise Pumain.
\newblock Scaling laws and urban systems.
\newblock \emph{Santa Fe Institute, Working Paper n 04-02}, 2:\penalty0 26,
  2004.

\bibitem[Bettencourt et~al.(2007)Bettencourt, Lobo, Helbing, K{\"u}hnert, and
  West]{Bettencourt:2007}
Lu{\'\i}s~MA Bettencourt, Jos{\'e} Lobo, Dirk Helbing, Christian K{\"u}hnert,
  and Geoffrey~B West.
\newblock Growth, innovation, scaling, and the pace of life in cities.
\newblock \emph{Proceedings of the National Academy of Sciences}, 104\penalty0
  (17):\penalty0 7301--7306, 2007.

\bibitem[Shigesada and Kawasaki(1997)]{Shigesada:1997}
N.~Shigesada and K.~Kawasaki.
\newblock \emph{Invasion by stratified diffusion}, chapter~5, page 79–103.
\newblock Oxford University Press, USA, 1997.

\bibitem[Clark et~al.(2001)Clark, Lewis, and L.]{Clark:2001}
J.S. Clark, M.~Lewis, and Horvath L.
\newblock Invasion by extremes: Population spread with variation in dispersal
  and reproduction.
\newblock \emph{The American Naturalist}, 157\penalty0 (5):\penalty0 537--554,
  2001.

\bibitem[Iwata et~al.(2000)Iwata, Kawasaki, and Shigesada]{Iwata:2000}
K.~Iwata, K.~Kawasaki, and N.~Shigesada.
\newblock A dynamical model for the growth and size distribution of multiple
  metastatic tumors.
\newblock \emph{J. Theor. Biol.}, 203\penalty0 (2):\penalty0 177--186, 2000.

\bibitem[Haustein and Schumacher(2012)]{Haustein:2012}
V~Haustein and U~Schumacher.
\newblock A dynamical model for tumour growth and metastasis formation.
\newblock \emph{J. Clin. Bioinforma.}, 2\penalty0 (1), 2012.

\bibitem[Fisher(1937)]{Fisher:1937}
Ronald~Aylmer Fisher.
\newblock The wave of advance of advantageous genes.
\newblock \emph{Annals of Human Genetics}, 7\penalty0 (4):\penalty0 355--369,
  1937.

\bibitem[Shigesada and Kawasaki(2002)]{Shigesada:2002}
N.~Shigesada and K.~Kawasaki.
\newblock \emph{Invasion and the range expansion of species: effects of
  long-distance dispersal}, chapter~17, page 350–373.
\newblock Blackwell Science, 2002.

\bibitem[Stanilov(2013)]{Stanilov:2013}
Kiril Stanilov.
\newblock Planning the growth of a metropolis: factors influencing development
  patterns in west london, 1875--2005.
\newblock \emph{Journal of Planning History}, 12\penalty0 (1):\penalty0 28--48,
  2013.

\bibitem[Carra et~al.(2017)Carra, Mallick, and Barthelemy]{Carra:2017}
Giulia Carra, Kirone Mallick, and Marc Barthelemy.
\newblock Coalescing colony model: Mean-field, scaling, and geometry.
\newblock \emph{Physical Review E}, 96\penalty0 (6):\penalty0 062316, 2017.

\bibitem[Bouchaud(2019)]{Bouchaud:2019}
Jean-Philippe Bouchaud.
\newblock Econophysics: Still fringe after 30 years?
\newblock \emph{arXiv preprint arXiv:1901.03691}, 2019.

\bibitem[Zipf(1949)]{Zipf:1949}
George~Kingsley Zipf.
\newblock \emph{Human behavior and the principle of least effort.}
\newblock addison-wesley press, 1949.

\bibitem[Batty(2006)]{Batty:2006}
Michael Batty.
\newblock Rank clocks.
\newblock \emph{Nature}, 444\penalty0 (7119):\penalty0 592--596, 2006.

\bibitem[Soo(2005)]{Soo:2005}
Kwok~Tong Soo.
\newblock Zipf's law for cities: a cross-country investigation.
\newblock \emph{Regional science and urban Economics}, 35\penalty0
  (3):\penalty0 239--263, 2005.

\bibitem[Gibrat(1931)]{Gibrat:1931}
Robert Gibrat.
\newblock \emph{Les in{\'e}galit{\'e}s {\'e}conomiques}.
\newblock Recueil Sirey, 1931.

\bibitem[Marsili and Zhang(1998)]{Marsili:1998}
Matteo Marsili and Yi-Cheng Zhang.
\newblock Interacting individuals leading to zipf's law.
\newblock \emph{Physical Review Letters}, 80\penalty0 (12):\penalty0 2741,
  1998.

\bibitem[Gabaix(1999)]{Gabaix:1999}
Xavier Gabaix.
\newblock Zipf's law for cities: an explanation.
\newblock \emph{Quarterly journal of Economics}, pages 739--767, 1999.

\bibitem[Sornette and Cont(1997)]{Sornette:1997}
Didier Sornette and Rama Cont.
\newblock Convergent multiplicative processes repelled from zero: power laws
  and truncated power laws.
\newblock \emph{Journal de Physique I}, 7\penalty0 (3):\penalty0 431--444,
  1997.

\bibitem[Bouchaud and M{\'e}zard(2000)]{Bouchaud:2000}
Jean-Philippe Bouchaud and Marc M{\'e}zard.
\newblock Wealth condensation in a simple model of economy.
\newblock \emph{Physica A: Statistical Mechanics and its Applications},
  282\penalty0 (3):\penalty0 536--545, 2000.

\bibitem[Brueckner(1987)]{Brueckner:1987}
Jan~K Brueckner.
\newblock The structure of urban equilibria: A unified treatment of the
  muth-mills model.
\newblock \emph{Handbook of regional and urban economics}, 2:\penalty0
  821--845, 1987.

\bibitem[Glaeser et~al.(2008)Glaeser, Kahn, and Rappaport]{Glaeser:2008}
Edward~L Glaeser, Matthew~E Kahn, and Jordan Rappaport.
\newblock Why do the poor live in cities? the role of public transportation.
\newblock \emph{Journal of urban Economics}, 63\penalty0 (1):\penalty0 1--24,
  2008.

\bibitem[Krugman(1996)]{Krugman:1996}
Paul~R Krugman.
\newblock \emph{The self-organizing economy}.
\newblock Blackwell Oxford, 1996.

\bibitem[Louf and Barthelemy(2013)]{Louf:2013}
R{\'e}mi Louf and Marc Barthelemy.
\newblock Modeling the polycentric transition of cities.
\newblock \emph{Physical Review Letters}, 111\penalty0 (19):\penalty0 198702,
  2013.

\bibitem[Louail et~al.(2014)Louail, Lenormand, Ros, Picornell, Herranz,
  Frias-Martinez, Ramasco, and Barthelemy]{Louail:2014}
Thomas Louail, Maxime Lenormand, Oliva G~Cantu Ros, Miguel Picornell, Ricardo
  Herranz, Enrique Frias-Martinez, Jos{\'e}~J Ramasco, and Marc Barthelemy.
\newblock From mobile phone data to the spatial structure of cities.
\newblock \emph{Scientific reports}, 4, 2014.

\bibitem[Dyson(1962)]{Dyson:1962}
Freeman~J Dyson.
\newblock Statistical theory of the energy levels of complex systems. i.
\newblock \emph{Journal of Mathematical Physics}, 3\penalty0 (1):\penalty0
  140--156, 1962.

\bibitem[Branston(1976)]{Branston:1976}
David Branston.
\newblock Link capacity functions: A review.
\newblock \emph{Transportation Research}, 10\penalty0 (4):\penalty0 223--236,
  1976.

\bibitem[Louf and Barthelemy(2014)]{Louf:2014}
R{\'e}mi Louf and Marc Barthelemy.
\newblock How congestion shapes cities: from mobility patterns to scaling.
\newblock \emph{Scientific Reports}, 4, 2014.

\bibitem[Verbavatz and Barthelemy(2019)]{Verbavatz:2019}
Vincent Verbavatz and Marc Barthelemy.
\newblock Critical factors for mitigating car traffic in cities.
\newblock \emph{arXiv preprint arXiv:1901.01386}, 2019.

\bibitem[Depersin and Barthelemy(2018)]{Depersin:2018}
Jules Depersin and Marc Barthelemy.
\newblock From global scaling to the dynamics of individual cities.
\newblock \emph{Proceedings of the National Academy of Sciences}, 115\penalty0
  (10):\penalty0 2317--2322, 2018.

\bibitem[Chang et~al.(2017)Chang, Lee, and Choi]{Chang:2017}
Yu~Sang Chang, Yong~Joo Lee, and Sung Sup~Brian Choi.
\newblock Is there more traffic congestion in larger cities?-scaling analysis
  of the 101 largest us urban centers.
\newblock \emph{Transport Policy}, 59:\penalty0 54--63, 2017.

\bibitem[Bouchaud et~al.(1998)Bouchaud, Cugliandolo, Kurchan, and
  Mezard]{Bouchaud:1998}
Jean-Philippe Bouchaud, Leticia~F Cugliandolo, Jorge Kurchan, and Marc Mezard.
\newblock Out of equilibrium dynamics in spin-glasses and other glassy systems.
\newblock \emph{Spin glasses and random fields}, pages 161--223, 1998.

\bibitem[Strano et~al.(2012)Strano, Nicosia, Latora, Porta, and
  Barthelemy]{Strano:2012}
Emanuele Strano, Vincenzo Nicosia, Vito Latora, Sergio Porta, and Marc
  Barthelemy.
\newblock Elementary processes governing the evolution of road networks.
\newblock \emph{Scientific reports}, 2, 2012.

\bibitem[Levinson and Yerra(2006)]{Levinson:2006}
David Levinson and Bhanu Yerra.
\newblock Self-organization of surface transportation networks.
\newblock \emph{Transportation Science}, 40\penalty0 (2):\penalty0 179--188,
  2006.

\bibitem[Kivel{\"a} et~al.(2014)Kivel{\"a}, Arenas, Barthelemy, Gleeson,
  Moreno, and Porter]{Kivela:2014}
Mikko Kivel{\"a}, Alex Arenas, Marc Barthelemy, James~P Gleeson, Yamir Moreno,
  and Mason~A Porter.
\newblock Multilayer networks.
\newblock \emph{Journal of Complex Networks}, 2\penalty0 (3):\penalty0
  203--271, 2014.

\bibitem[Gallotti and Barthelemy(2014)]{Gallotti:2014}
Riccardo Gallotti and Marc Barthelemy.
\newblock Anatomy and efficiency of urban multimodal mobility.
\newblock \emph{Scientific reports}, 4, 2014.

\bibitem[Strano et~al.(2015)Strano, Shai, Dobson, and Barthelemy]{Strano:2015}
Emanuele Strano, Saray Shai, Simon Dobson, and Marc Barthelemy.
\newblock Multiplex networks in metropolitan areas: generic features and local
  effects.
\newblock \emph{Journal of The Royal Society Interface}, 12\penalty0
  (111):\penalty0 20150651, 2015.

\bibitem[Sobstyl et~al.(2018)Sobstyl, Emig, Qomi, Ulm, and
  Pellenq]{Sobstyl:2018}
JM~Sobstyl, T~Emig, MJ~Abdolhosseini Qomi, F-J Ulm, and RJ-M Pellenq.
\newblock Role of city texture in urban heat islands at nighttime.
\newblock \emph{Physical Review Letters}, 120\penalty0 (10):\penalty0 108701,
  2018.

\end{thebibliography}

\end{document}